# 3DMASC: ACCESSIBLE, EXPLAINABLE 3D POINT CLOUDS CLASSIFICATION. APPLICATION TO BI-SPECTRAL TOPO-BATHYMETRIC LIDAR DATA.

Mathilde Letard[1*], Dimitri Lague[1,2*], Arthur Le Guennec[1,3], Sébastien Lefevre[4], Baptiste Feldmann[1,2], Paul Leroy[1,2], Daniel Girardeau-Montaut[5] and Thomas Corpetti[3]

[1] Univ Rennes, Geosciences Rennes, UMR 6118 CNRS, France.
[2] Univ Rennes, Plateforme LiDAR Topo-Bathymétrique Nantes-Rennes, OSUR, UAR 3343 CNRS, France.
[3] LETG UMR 6554, CNRS, F-35000 Rennes, France
[4] IRISA UMR 6074, Université Bretagne Sud, F-56000 Vannes, France
[5] Johnson and Johnson
* Correspondence: dimitri.lague@univ-rennes1.fr, mathilde.letard@univ-rennes1.fr

**ABSTRACT:**

Three-dimensional data have become increasingly present in earth observation over the last decades. However, many 3D surveys are still underexploited due to the lack of accessible and explainable automatic classification methods, for example, new topo-bathymetric lidar data. In this work, we introduce explainable machine learning for 3D data classification using Multiple Attributes, Scales, and Clouds under 3DMASC, a new workflow. This workflow introduces multi-cloud classification through dual-cloud features, encrypting local spectral and geometrical ratios and differences. 3DMASC uses classical multi-scale descriptors adapted to all types of 3D point clouds and new ones based on their spatial variations. In this paper, we present the performances of 3DMASC for multi-class classification of topo-bathymetric lidar data in coastal and fluvial environments. We show how multivariate and embedded feature selection allows the building of optimized predictor sets of reduced complexity, and we identify features particularly relevant for coastal and riverine scene descriptions. Our results show the importance of dual-cloud features, lidar return-based attributes averaged over specific scales, and of statistics of dimensionality-based and spectral features. Additionally, they indicate that small to medium spherical neighbourhood diameters (<7 m) are sufficient to build effective classifiers, namely when combined with distance-to-ground or distance-to-water-surface features. Without using optional RGB information, and with a maximum of 37 descriptors, we obtain classification accuracies between 91% for complex multi-class tasks and 98% for lower-level processing using models trained on less than 2000 samples per class. Comparisons with classical point cloud classification methods show that 3DMASC features have a significantly improved descriptive power. Our contributions are made available through a plugin in the CloudCompare software, allowing non-specialist users to create classifiers for any type of 3D data characterized by 1 or 2 point clouds (airborne or terrestrial lidar, structure from motion), and two labelled topo-bathymetric lidar datasets, available on https://opentopography.org/.

## 1. INTRODUCTION

3D data are becoming increasingly popular among geoscientists, as they constitute major opportunities for enhanced observation of natural processes and more precise risk assessments. In particular, in complex, natural environments combining vegetated terrains, artificialized portions, and submerged areas, specific 3D point clouds obtained through topo-bathymetric (TB) lidar sensors are an opportunity to gather knowledge inaccessible to other surveying methods. Indeed, TB lidar sensors were introduced specifically to enable the documentation of shallow waters at high resolution (Fernandez-Diaz et al., 2016, 2014; Mandlburger et al., 2015; McKean et al., 2009; Quadros et al., 2008; Wang and Philpot, 2007). These sensors combine the strengths of two types of lidar sensors. First, topographic lidars with small footprint near-infrared (NIR) lasers and high shot densities that cannot penetrate water. Second, large footprint bathymetric lidars, able to image seafloors deeper than 30 m in clear waters (Guenther et al., 2000; Philpot, 2019), but with reduced point density and spatial resolution, and high mobilization costs. TB lidar sensors practically combine both types of sensors; a NIR laser ($\lambda$=1064 nm) and a green laser ($\lambda$=532 nm), and their respective benefits. Associated TB lidar datasets are bi-spectral, consisting of one point cloud (PC) per wavelength, with submerged topographies as detailed as emerged parts (see Figure 1). Due to the different specificities of each laser, namely their footprint size, scanning angle range, and wavelength, the obtained PCs are systematically different and provide distinct samplings of the same scene, particularly over vegetation and submerged surfaces. TB lidar sensors are useful in the study of varying subjects. Combining high-resolution data about the submerged and emerged surfaces offers new opportunities to map habitats in fluvial (Fernandez-Diaz et al., 2014; Mandlburger et al., 2015; McKean et al., 2009; Pan et al., 2015) or coastal (Chust et al., 2010; Hansen et al., 2021; Launeau et al., 2018; Parrish et al., 2016; Smeeckaert et al., 2013; Wilson et al., 2019) environments, improve high-resolution modelling of flood inundation (Lague and Feldmann, 2020; Mandlburger et al., 2015) or track sediment transport at the land-water interface. These bispectral sensors have also been shown to be useful for vegetated or urbanized terrain assessment (Dai et al., 2018; Ekhtari et al., 2018; Laslier et al., 2019; Morsy et al., 2017b; Wichmann et al., 2015). However, adapted processing methods are necessary to fully use them and leverage the scientific potential of such extensive datasets over complex natural scenes. In particular, automatic classification of the green lidar data directly at the 3D PC level is essential.

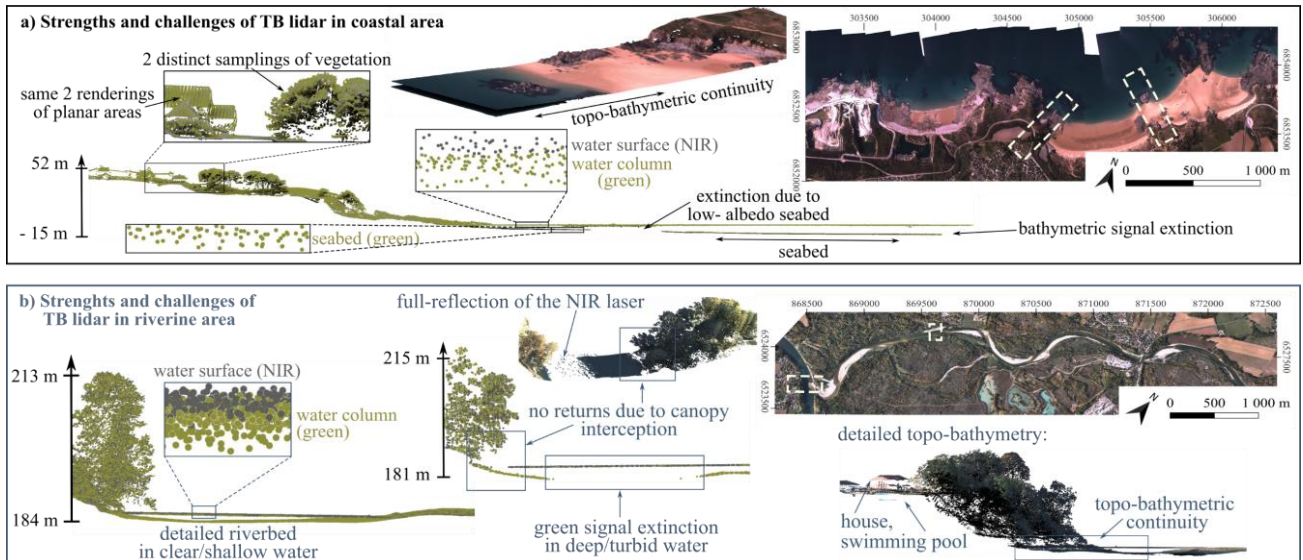

Figure 1: Strengths and challenges of topo-bathymetric lidar data. Examples of (a) the coastal setting of the surroundings of Fréhel (France) and (b) along the Ain River (France). Datasets are presented in the RGF93 coordinates system.

In such areas, the exploitation of the PCs obtained with both wavelengths is beneficial, mainly (i) because the rich spectral information provided by the combined surveys can be leveraged to distinguish vegetated, submerged, and urban objects that can be mixed in the same scenes, and (ii) because the increased geometric information provided by a simultaneous sampling with two lasers provides additional information and chances to image a greater portion of the terrain.

Several methodological challenges complexify the development of adapted classification workflows. First, the refraction of the green laser in water makes it critical to detect all green points below water during data production to subsequently perform accurate refraction correction on the position of the received echoes. This correction requires accurately knowing the spatial extent of water in the scene and the local water elevation, which the NIR channel gives when data are available in the area. While relatively straightforward in coastal environments or large lakes as water has a constant elevation, detecting water surfaces is far more challenging in fluvial environments for four reasons: (i) water elevation decreases downstream, sometimes abruptly at the vicinity of dams; (ii) rivers can have several active braids or complex hydrological connection with abandoned channels or lakes in adjacent floodplains; (iii) full mirror-like NIR reflection may occur on flat water such that the NIR PC may lack water surface echoes over large areas (Figure 1); (iv) vegetation frequently grows on the floodplain such that river banks and small lakes may be completely below vegetation making things even more complex as canopy interception reduces the backscattered intensity and the likelihood of having a water surface NIR echo and bottom green echo (Figure 1).

Second, the backscattered green laser energy generates two prominent echoes in an ideal clear water column. The first is a volume echo located just below the water surface. Its position can deviate from the actual water surface from several dozens of centimetres depending on water characteristics, leaving no other possibility than to use the corresponding NIR survey to derive the real water surface.(Guenther et al., 2000;Lague and Feldmann, 2020; Philpot, 2019). Though this volume echo is of no use, it is systematic for any shot. The second echo corresponds to the bathymetry. However, in turbid or deep waters, it sometimes has such a weak amplitude that its signal-to-noise ratio hinders its detection. For a given sensor and flight elevation, the maximum measurable water depth thus highly depends on water clarity and bottom reflectance (Guenther et al., 2000; Lague and Feldmann, 2020; Philpot, 2019). For instance, in clear coastal waters, the Teledyne Optech Titan sensor can reach depth down to 10-15 m over bright sand but can be limited to 0.5 m over dark rocks and will typically only reach depths of 1-4 m in rivers owing to the reduced water clarity (Lague and Feldmann, 2020). Thus, it is commonplace in inland water surveys that deeper parts of rivers or lakes are locally not detected due to green laser extinction.

Consequently, as for ground detection below dense vegetation, one cannot assume that simple operations such as picking the lowest green point over a specific area or extracting the last recorded echo in the green PC will systematically isolate the seabed or riverbed. Similarly, because of the green water surface uncertainty and the incomplete sampling of the NIR water surface, removing all green points below a given depth is impossible as a large part of the very shallow seabed will be discarded, and depth may be mis-estimated.

While the use of both NIR and green PCs appears essential to derive an accurate classification of submerged parts, the exact method is not straightforward, and there is currently no available solution to separate bathymetric echoes from volume automatically over large PC datasets in complex inland water environments.

Additionally, beyond detecting and separating bathymetric and volume echoes of the green laser, classifying the nature of the land-water continuum – seabed or riverbed covers and above-ground features – on 3D PCs is a significant challenge. Most of the existing approaches rely on 2D rasters classified with traditional algorithms like maximum likelihood, support vector machine, or decision trees (Letard et al., 2021; Sun and Shyue, 2017; Tulldahl and Wikström, 2012; Wedding et al., 2008; Zavalas et al., 2014). Although these methods exploit geometrical features, they analyse averaged features due to the rasterization step, which may produce mixed pixels (Hsieh et al., 2001) and smooth out the geometry of the scene, as the spatial point pattern information is lost when the data is condensed into regularly spaced observations. Few studies provide 3D classifications of underwater environments using bathymetric lidar (Hansen et al., 2021; Letard et al., 2022; Letard et al., 2022). Additionally, some approaches require full-waveform data (Letard et al., 2022; Letard et al., 2022), which is complex to process and often unavailable or unpublished. Over land, the bi-spectral backscattered intensity of TB lidar offers new classification opportunities, as explored in urban environments by Morsy et al., (2022, 2017a) and Teo and Wu (2017). The two distinct samplings of each laser, provided by their different footprint sizes, can also potentially provide useful information. However, there have been, to date, no applications attempting to classify both clouds of a topobathymetric survey directly. Fusing them into a single cloud to apply workflows existing for forested or urban environments is impossible as mixing the data obtained with the two different sensors would be incorrect due to their different optical and echoes characteristics. Spectral and multi-echo-based information would become unusable, and the water surface sampling obtained would be unexploitable due to the reasons explained above. Additionally, the capacity of one or the other sensor to image specific parts of natural scenes is information in itself that would be lost if both clouds were to be fused for processing, while it can be exploited by directly operating on the differences between the point clouds. A processing method adapted to multiple clouds and applicable to configurations in which vegetated, urban, and submerged settings are combined is thus expected.

This work presents an original framework called 3DMASC for 3D point classification with Multiple Attributes, Multiple Scales, and Multiple Clouds and its application to coastal and fluvial TB airborne lidar datasets. 3DMASC operates directly at the 3D PC level to produce outputs in 3D, thus preserving the rich information of spatial point patterns. This classification process relies on multiple 3D features that make it generalisable to various 3D data types and point classes. By simultaneously assessing point cloud characteristics at different scales, it can distinguish classes characterized by different sizes while balancing salt and pepper-like noise or errors at the borders between classes that often come with small and large scales, respectively. Finally, our workflow operates directly on the differences between distinct samplings offered by multiple point clouds, thus leveraging the underexploited knowledge of multi-cloud surveys. 3DMASC combines proven classical elements of single PC semantic classification, such as geometric feature extraction from multi-scale spherical neighbourhoods (Brodu and Lague, 2012) or k-nearest neighbours (Thomas et al., 2018) and a random forest model (Breiman, 2001). In addition, it adds new features specifically engineered to leverage the NIR and green PCs. Our contributions consist of the following:

- Designing new joint-cloud features calculated on two PCs using their local geometry and backscattered intensity. 3DMASC uses a flexible method to compute features from two PCs, resulting in more than 80 different features;
- Screening over 80 features, both classical and new, to select the essential features and scales contributing to 3D point classification to optimize classifiers in terms of computational efficiency, generalization ability, and interpretability;
- Demonstrating that with limited training data (< 2000 points per class) and less than ten features and five scales, the classification accuracy of TB lidar datasets can be excellent (>0.95);
- Providing a plugin in the open source software Cloudcompare (CC) that can be used easily by non-specialists to classify any 3D PC and by experts for fast 3D feature computation and visualization;
- Sharing two manually labelled state-of-the-art lidar datasets with two different levels of detail (up to 13 classes).

The paper is organized as follows: the next section introduces the related works on the processing of 3D point clouds; our methodology is then introduced in sections 3 and 4, and experimental results are shown in section 5, associated with a discussion in section 6, and a conclusion in section 7.

## 2. RELATED WORK

Classification of 3D data is a challenge, as 3D data are unstructured, irregular, and unordered. Their characterization is made harder by the local density variations and the complex objects they contain. In this section, we review methods producing supervised 3D classifications. Clustering methods and approaches relying on rasterized lidar data are not reviewed. Existing supervised 3D point cloud classification methods can be organized into two categories: handcrafted features with conventional classifiers and learned features with deep neural networks.

### 2.1 Learned features with neural networks

**Deep neural networks** consist of interconnections of neurons organized in layers. Each neuron performs a linear combination of its inputs associated with an activation function. The connection of a potentially large number of neurons, organized depending on applications (the so-called architecture), enables the modelling of very complex functions. The training stage, performed by backpropagation, consists of estimating parameters for each neuron. Neural networks can adapt and generalize their learning to new inputs, making them powerful tools in scientific research and machine learning applications. Through this process, features of the data progressively stand out and are used to build task-adapted prediction rules. Neural networks thus learn relevant features directly from the data and eliminate the need to define features and scales upstream, contrary to classical machine learning approaches.

In general, neural networks are mainly based on linear combinations or convolution operators. When 3D data emerged, their processing with neural networks posed many conceptual and computational issues. One first issue is the heavy computations needed to load and process data in three dimensions. A second major issue was to adapt proven methods to unstructured, irregular 3D data. As a result, although convolutional neural networks (CNN) are among the most performant deep neural networks, partly thanks to their ability to extract high-level features while considering spatial context, their application to 3D PCs is not straightforward. Their development is thus recent and includes a wide range of variations to optimize performances and complexity.

Some networks perform convolution using transformed points (Atzmon et al., 2018; Hua et al., 2018; Li et al., 2018; Xu et al., 2018), voxelized PCs (Graham et al., 2018; Tchapmi et al., 2018), graphs obtained from adjacencies between points (Mao et al., 2022c; L. Wang et al., 2019; Y. Wang et al., 2019; Wei et al., 2023; Wen et al., 2021) or groups of points (Hui et al., 2021; Landrieu and Boussaha, 2019; Landrieu and Simonovsky, 2018) – called superpoints –, or kernel points (Thomas et al., 2019).

Currently, KPConv (Thomas et al., 2019) and SPG (Landrieu and Simonovsky, 2018) are among the state-of-the-art architectures for 3D point classification. However, research around 3D deep learning is very active and is working towards different improvements of existing solutions. Examples of recent experiments include spatially sparse convolutions (Graham et al., 2018; Schmohl and Sörgel, 2019) specifically designed to handle the sparsity of PCs. Some architectures also incorporate state-of-the-art 2D deep learning solutions to 3D networks, such as attention mechanisms (Deng and Dong, 2021; Huang et al., 2021; L. Wang et al., 2019; Zeng et al., 2023; Zhang et al., 2022) or residual connections (Huang et al., 2018; Ye et al., 2018; Zeng et al., 2023). In a similar attempt to adapt 2D principles to 3D, multiple projects combine several receptive fields as in inception networks to improve 3D processing (Mao et al., 2022a, 2022c). 3D deep neural networks have also started being applied to airborne lidar surveys (Huang et al., 2021; Lin et al., 2021; Mao et al., 2022a; Schmohl and Sörgel, 2019; Wen et al., 2021; Yang et al., 2018; Zeng et al., 2023; Zhang et al., 2022; Zhao et al., 2018). However, existing work on airborne lidar processing with 3D deep neural networks exclusively concerns urban areas or forests and thus does not address the issue of complex and diverse natural or semi-urban environments.

Current evolutions also tackle the issue of computational cost and complexity of 3D deep neural networks. Uses of the Transformer architecture on 3D data limit the amount of power needed to compute convolutions (Cheng et al., 2023; M. H. Guo et al., 2021; Lai et al., 2022; Park et al., 2022; Robert et al., 2023; H. Zhao et al., 2021), while networks performing on features rather than points offer faster, lighter alternatives (Gao et al., 2023). To simplify the use of 3D deep neural networks, approaches to train the algorithms on a limited amount of labelled data currently arise. An option is to rely on few-shot learning, thus drastically reducing the number of labelled samples required to train (Dong and Xing, 2018; Feng et al., 2022; Garcia and Bruna, 2018; He et al., 2023; Li et al., 2022; Mao et al., 2022b; Xu and Lee, 2020; N. Zhao et al., 2021). Domain adaptation can also transfer previously learned weights to unlabelled data (Cazorla et al., 2022; Jaritz et al., 2023; Yuan et al., 2023).

Nevertheless, these developments are still recent, and although highly performing, 3D deep neural networks are still experimental and less accessible to non-specialist users than more classical machine learning algorithms. They require dedicated and well-parameterized graphics processing units (GPUs), which limits their accessibility to a wide range of environmental researchers. Their increased complexity to handle, optimize and parameterize without extended knowledge about machine learning, scientific programming and mathematics is another factor limiting their deployments for 3D analyses by thematic users. The recent advances towards lower supervision and smaller amounts of labelled data still show lower performances than fully-supervised approaches. Finally, to our knowledge, there are no methods capable of considering two point clouds derived from two different wavelengths. As a consequence, because of the lack of foundation models for 3D data processing in natural environments, the low availability of labelled data in such scenes, and the more complex and debated explainability of deep learning models, we prefer to turn towards machine learning on handcrafted features, as our goal was to provide a generalisable and accessible tool.

**2.2 Handcrafted features and conventional classifiers**

In this section, we review the methodological context on which we based our framework and introduce the different concepts we incorporate or start from to build 3DMASC.

2.2.1 Features definition

Supervised machine learning classifiers require the definition of an input vector to feed to the classifier. This vector often consists of a group of handcrafted data attributes – also called features or descriptors – that encode characteristics of the points and their context. The spatial repartition of the points in a PC and, for multiple return lidar, the echoes' number, ordering and characteristics depend on a combination of sensor physics and surface geometry. The reflected intensity is also linked to the albedo of the surveyed object and to the sensor. They thus act as proxies of the actual surface characteristics.

PC classifications consequently exploit the **geometry of the PCs** (Hackel et al., 2016) and their spectral dynamics (Chehata et al., 2009) or their **local dimensionality** (Brodu and Lague, 2012; Vandapel et al., 2004). For example, the eigenvectors of each point's neighbourhood covariance matrix are popular attributes in identifying isolated points, lines, planes, volumes, contours and edges in PCs (Gross and Thoennessen, 2006). Using ratios of these eigenvalues allows to assess the linearity, planarity, sphericity, anisotropy, eigenentropy, omnivariance, scattering or change of curvature of the 3D shape (Chehata et al., 2009; Gross and Thoennessen, 2006; Pauly, 2003). Estimates of the PC's local point density (Weinmann et al., 2013) or the verticality (Demantké et al., 2012) are other helpful parameters to classify points. **Multiples return characteristics** associated with airborne lidar data also constitute information on the objects surveyed: the number of returns, return number or ratio of both are useful for identifying ground, buildings or vegetation (Chehata et al., 2009). **Height-derived features** such as elevation variations between points of a neighbourhood or point distribution kurtosis or skewness are also used for classification purposes (Antonarakis et al., 2008; Chehata et al., 2009; Guan et al., 2012; Yan et al., 2015). They can be combined with **distance-to-ground features**, corresponding to the distance between the points to classify and the ground, through the analysis of a digital terrain model, for example (Blomley and Weinmann, 2017; Chehata et al., 2009; Niemeyer et al., 2012).

Another possibility to assess the geometrical characteristics of point clouds locally is to use **histogram-based features** that analyse the variations of geometrical features around the point through histograms, whose bins are used as descriptors (Blomley et al., 2016; Blomley and Weinmann, 2017; Himmelsbach et al., 2009; Osada et al., 2002; Rusu et al., 2009; Tombari et al., 2010; Wohlkinger and Vincze, 2011). However, computing these features experimentally demands a higher computation time (Garstka and Peters, 2016). We thus did not include them in the proposed method but compared their performance to our results in Section 6.

Though some studies solely exploit PC geometry to identify 3D objects (West et al., 2004), the **radiometric information** contained in lidar data can further improve PC interpretation where objects have similar geometries (Yan et al., 2015). Radiometric information is rarely used on its own (Song et al., 2002) and is often integrated as a complement to previously mentioned geometrical features. It is often among the most contributive features to improve segmentation (Dai et al., 2018) and classification results (Im et al., 2008). The most popular attribute is the mean value of the backscattered intensity over a neighbourhood or between the first and last returns (Antonarakis et al., 2008). Combining **multispectral radiometric measurements** provides even more reliable information than single wavelength data (Morsy et al., 2017b). Using multispectral lidar systems allows to incorporate intensity ratios – for example, vegetation indexes – to classification predictors (Chen et al., 2017; Morsy et al., 2017b; Wichmann et al., 2015) or to compare surface reflectances in different optical domains (Chen et al., 2017; Gong et al., 2015) and even create colour composites with different channel combinations (Wichmann et al., 2015), thus refining point identification (Im et al., 2008). However, these existing methods perform nearest neighbour interpolations of different intensities. Among the reviewed features, no solution to formalize the sampling differences of multi-spectral point clouds exists, so we aim to build novel multi-cloud descriptors.

2.2.2 Features extraction

3D point clouds are unordered and have varying densities. Also, points are 0-dimensional and thus do not contain any meaningful geometrical information other than their Z coordinate. For these reasons, descriptors are computed on the **neighbourhood** of each point, which describes the local geometry. The spherical neighbourhood is the most common for PC processing, defined by its radius or diameter and comprising each point's nearest neighbours with respect to 3D Euclidean distance. Cylindrical neighbourhoods are also exploited by Niemeyer et al. (2012), and cubic or cuboid ones are explored by Dong et al., (2017). Overall, spherical neighbourhoods are considered the most helpful, based on the observations of Thomas et al. (2018) and Hermosilla et al., (2018), which compared the use of nearest neighbours (NN) and spherical searches to describe PCs. They are considered more stable than NN to the variations of density (Hermosilla et al., 2018), surface slope or orientation and point pattern that occur in PCs, and more efficient for handcrafted feature extraction (Thomas et al., 2018). Thomas et al. (2019) additionally states that a consistent spherical domain helps classifiers learn more meaningful representations of the local aspect of the PCs during training. Based on these observations, we mainly exploit spherical neighborhoods in this work.

Independently from the type of neighbourhood implemented, descriptive features of 3D data can be computed at a **single constant scale** (Chehata et al., 2009) or **multiple scales** (Blomley and Weinmann, 2017; Brodu and Lague, 2012; Hackel et al., 2017, 2016; Niemeyer et al., 2012). Multiple scales successively applied to each point have proven to have greater descriptive power than a single constant scale since they can better capture scene elements of different sizes (e.g., vegetation) and the variations of object geometry with scale (Brodu and Lague, 2012; Hackel et al., 2017, 2016; Thomas et al., 2018). Considering the diversity of objects in PCs, the neighbourhood type, the number of scales used, and their values impact the classification of the data and thus require careful parameterization. Automatic **optimal scale identification** has been investigated to avoid empiric selection. It mainly relies on minimizing information redundancy – through correlation or entropy estimates – and maximizing relevance in terms of classification accuracy. For single-scale classification, Niemeyer et al. (2011) advised an optimal scale of 7 NN in terms of classification accuracy when classifying urban scenes with lidar data. Rather than defining a fixed set of multiple scales, Demantke et al. (2011) try to identify automatically the most relevant scale to describe each point's neighbourhood by using its dimensionality. Similarly, Weinmann et al. (2015) select each point's individual optimal scale before extracting and selecting descriptive features. These approaches combine the use of multiple scales across the PC and the computation of features at a single scale for each point. Dong et al. (2017) propose to select an optimal neighbourhood type and its scale for each feature rather than optimizing the scale for each point, thus combining the advantages of different types of neighbourhoods, multiple scales and uncorrelated features. Our framework builds on these propositions to exploit the strengths of multi-scale neighborhoods while limiting computational complexity.

2.2.3 Features selection

The number of attributes derived from the data to learn the prediction rule may be large. However, introducing a wide range of information to represent the data may not always be optimal. Some of the multiple descriptive features may be irrelevant regarding the target variable or redundant. Although irrelevant or redundant information should theoretically not impact the prediction, it can impair the learning process by increasing the number of parameters involved and complexifying the optimization process. Similarly to the scales exploited, an optimized feature set should thus incorporate the most information possible while also limiting redundancy between attributes. Considering the variety of information derivable from 3D data, empirically selecting the attributes to integrate into a classification is time-consuming and hazardous, as it might impair classification performances. **Feature selection** methods allow the automation of a great part of the feature vector construction. They are mainly based on estimating an attribute's relevance relatively to the predicted variables and minimising the correlation between relevant parameters. As explained in Dash and Liu, (1997), feature selection methods can be split into three categories. **Filter-based or univariate methods** aim at maximizing the relevance of the predictors used. They use relevance score functions and rankings of the scores to keep only a subset of the most informative features for classification. Popular score functions include Fisher's index or Information Gain index but adapted metrics that consider multiple aspects of feature relevance also exist (Weinmann et al., 2013). **Multivariate methods** try to minimize feature redundancy among the relevant attributes, often by combining score functions with correlation assessments (Dong et al., 2017; Weinmann et al., 2015). Both univariate and multivariate approaches are independent of the classifier used and its settings, which is sometimes seen as a generalization advantage (Weinmann et al., 2015), but also do not account for inter-feature synergies, and may evict highly correlated but still informative features (Guyon and Elisseeff, 2003). **Wrapper methods and embedded feature selection** consist of exploiting classifier outputs to select features. They either use classification accuracy obtained using each feature separately as a score to prune the input vector (Dong et al., 2017) through backwards or forward selection or rely on feature importance information provided by algorithms, to evict the least important predictors and improve accuracy (Guan et al., 2012). Random forest-based metrics are among the most common embedded selection strategies. To limit the complexity of our classifiers, we rely on multi-variate and embedded feature selection strategies to prune our feature vectors.

2.2.4 3D Points classification

Many classification algorithms have been developed to classify 3D PCs. The most common ones **classify each point individually** without considering the relationships between the point's label and its neighbour's assigned labels. They include instance-based techniques such as NN classification, rule-based predictions as applied by decision trees, probabilistic learners like Maximum Likelihood, max-margin learners as Support Vector Machines, and ensemble learning (Kotsiantis et al., 2007; Sagi and Rokach, 2018). Ensemble learning (Sagi and Rokach, 2018) is the most popular among individual point classification strategies. It relies on bagging, which consists of assembling several independent weak learners and combining them into a single strong learner using a voting mechanism. Random Forest (RF) models implement ensemble learning. Their ease of use, efficiency, robustness to overfitting, generalization abilities and production of a feature importance metric (Breiman, 2001; Pal, 2007) explain their frequent use for 3D data classification. They have been used for point-based classifications of both topographic and TB lidar (Chehata et al., 2009; Hansen et al., 2021; Letard et al., 2022; Letard et al., 2022). In RF, since the decision trees are independent, one cannot compensate for the potential weaknesses of another to improve the global performance of the forest. Algorithms like AdaBoost (Hastie et al., 2009) and XGBoost (Chen and Guestrin, 2016) overcome this limitation by incorporating boosting, which consists of training each weak learner to correct their predecessor's errors; however, they require more parameters compared to RF.

Individual point classifiers can only consider the spatial context of each point by encrypting it into the feature vector. However, they ignore that neighbour points' labels tend to be linked. Some algorithms thus implement **contextual classification,** which involves estimating the relationships between 3D points from a neighbourhood – often different from the one used for feature extraction – in the training data. Additionally to the classification performance objective, they aim to produce spatially consistent classifications of 3D PCs, avoiding the noisy output that individual point classifiers can produce. Thus, they tend to reach higher accuracies. Examples of such approaches are applications of Associative (Munoz et al., 2008; Triebel et al., 2006) and Non-associative Markov Networks (Najafi et al., 2014), Conditional Random Fields (Lim and Suter, 2009; Niemeyer et al., 2012, 2011b; Vosselman et al., 2017), and Markov Random Fields (Lu and Rasmussen, 2012) to 3D data. However, modelling 3D spatial relationships is computationally intensive and thus challenging to apply to large 3D datasets. These approaches also depend on the relationships observable in the training data, which makes exact inference of correlations between labels unattainable. In (Landrieu et al., 2017), a structured regularization framework allowing the conservation of a probabilistic approach and relying on a computationally lighter optimization is thus proposed, allowing the regularisation of any point-based classification with contextual information while keeping a form of precision information. In our work, we exploit the strengths of RF models and incorporate direct or indirect contextual knowledge through the descriptors used instead of exploiting regularisation frameworks.

**2.3 Classification of multispectral lidar point clouds**

TB lidars have the specificity of embedding two different non-co-focal lasers, thus producing two distinct samplings of the same scene under the form of two separate PCs whose points systematically have different positions. To our knowledge, no research on the application of 3D deep neural networks to TB areas surveyed with airborne bispectral lidar has yet been published. Existing approaches rely exclusively on handcrafted features extracted on full-waveforms (Launeau et al., 2018; Letard et al., 2022b), rasters (Wedding et al., 2008, Laslier et al., 2019), or PCs (Hansen et al., 2021). There is, however, more research on the classification of multispectral airborne lidar data over terrestrial areas using directly 3D PCs (Morsy et al., 2022, 2017b, 2017a, 2016, Wang and Gu, 2020), including an experimentation with PointNet++ over mixed urban areas (Jing et al., 2021). In this paper, however, the multispectral character of the data is summed up in one single point cloud with several intensity attributes, obtained using nearest neighbour interpolations over the three available spectral channels. There is thus no direct processing of multiple PCs simultaneously, as in Ekhtari et al., 2018, in which the authors create a synthetic individual multispectral point cloud from the three PCs of their multispectral survey and use it for classification. Other approaches exploit the information in each wavelength's point cloud individually and use it contiguously to perform classification (Letard et al., 2022b, Wang and Gu, 2020). Another existing possibility consists of computing spectral ratios between the different spectral channels to exploit the multiple wavelengths of the sensors (Matikainen et al., 2017; Morsy et al., 2022, 2017b, 2017a, 2016; Shaker et al., 2019). However, existing papers often do not include a wide range of possible classes, sometimes only tackling the problem of land/water distinction

(Shaker et al., 2019). Overall, there is a lack of available methods exploiting both the spectral and the geometrical differences between the different point clouds obtained through multispectral lidar surveys to classify land-water interface areas. In these areas, the sampling of the environment obtained with NIR or green wavelength is extremely distinct both spectrally and geometrically, providing key information about the environment (Figure 1). This paper addresses this research gap by proposing a multi-point cloud classification method.

## 3. FRAMEWORK/METHODOLOGY

In this section, we describe the 3DMASC method, included in a CloudCompare plugin (Girardeau-Montaut, 2022). The method consists first of computing descriptive features at multiple neighbourhoods of different sizes, involving one or two different PCs covering the area considered for classification. An optional features and scales selection method is then applied to reduce feature redundancy and ensure the relevance of the kept attributes. Finally, a random forest model is trained and used for classification, and its results are analyzed using prediction probability, Shapley explanations and feature importance values. Appendix A provides details of the implementation and operation.

### 3.1 3D features extraction

3DMASC operates directly on unordered sets of points, producing a 3D classification without requiring an intermediate rasterization step. A PC is a set of n 3D points {Pk | k=1,…,n} in which each element Pk is a vector of coordinates (x,y,z) with associated **point-based features**: *intensity*, *multi-echo characteristics* and potentially *red – green – blue (RGB) colour* (see Appendix B for a detailed list of features).

On top of point-based features, 3DMASC uses **neighbourhood-based** features defined using a spherical neighbourhood or a k-nearest neighbour search (kNN). A maximum of four different 3D entities are involved in the process of neighbourhood feature extraction:

- **1-2)** two point clouds. The originality of 3DMASC lies in using up to two PCs to characterize the scene of interest. For topo-bathymetric applications they originate from different wavelengths, typically 532 nm and 1064 nm. We refer to them as PC1 and PC2, respectively.
- **3)** A set of *core points* (Brodu and Lague, 2012), denoted PCX, that 3DMASC classifies at the end of the process. They may be a subset of points from PC1 or PC2 with a regular subsampling or other positions spread within the extent of PC1 and PC2.
- **4)** An optional *context* PC, denoted CTX, containing any relevant context information in its *Classification* attribute at a potentially much lower resolution than PC1 or PC2. A typical CTX would be previously classified ground points at 2 m spatial resolution.

### 3.1.1 Neighbourhood selection and scales

3DMASC mainly uses a spherical neighbourhood search in the relevant PC – PC1 or PC2, depending on the feature to compute – to capture the surroundings of each core point (Figure 2a). The neighbourhood scale is defined as the sphere diameter. 3DMASC uses a multi-scale classifier computing multiple neighbourhoods for each core point (Figure 2c). The user typically provides minimum and maximum scales and a step (e.g., 1 m) between successive scales. The minimum scale must be consistent with the PC's density to compute features for most core points. The size of the objects of interest typically sets the largest scale. Defining the optimal set of scales for various types of TB airborne lidar (e.g., coastal, fluvial…) is a challenge not yet resolved that we address in this work. Beyond ensuring classification success, it is crucial for operational efficiency, as the feature computation time increases strongly with the scale and the number of scales.

3DMASC also supports kNN to measure the vertical or horizontal distance between PC1 and PC2 or CTX (Figure 2b). This supplements relative position measurements between PCs where diameter-based features are impossible to compute due to a lack of neighbours.

### 3.1.2 Single cloud neighbourhood based features

Single cloud features describe PC1 or PC2 once at a time. Since many criteria characterize a 3D object and can help identify its nature, the plugin natively encompasses 15 features (see Appendix B for the complete list). The broad set of features available is presented in the following paragraphs.

**Six dimensionality-based features** aim to describe the local PC's general aspect and identify if the object has a linear, planar or spherical outlook (e.g., Brodu and Lague, 2012; Gross and Thoennessen, 2006; Vandapel et al., 2004). They rely on the eigenvalues of the covariance matrix of the points. 3DMASC can directly use the three normalized eigenvalues or classical combinations, resulting in sphericity, linearity and planarity metrics.

**Six geometry-based features** inform on the shape of the PC. 3DMASC computes and uses the slope angle, the detrended roughness, the curvature, the anisotropy, the number of points at a given scale and the first-order moment, introduced for contour detection in Hackel et al. (2017).

**Three height-based features** characterize the vertical structure of the local neighbourhood with respect to the minimum elevation $z_{min}$ and maximum elevation $z_{max}$. For a core point with elevation $z$, 3DMASC computes $z_{max}-z$, $z-z_{min}$ and the local thickness of the point cloud $z_{max}-z_{min}$, as explained in (Chehata et al., 2009).

**Optional contextual features** are used to place each core point in its spatial context and get its position relative to the ground, the water surface or any specific pre-existing class labelled in the CTX point cloud. They are computed with a kNN neighbourhood. They generalize the *distance to ground* feature used in Chehata et al. (2009) and Niemeyer et al. (2012). On top of these classical features, we propose novel 3D descriptors based on the application of **statistical operators on point-based features within spherical neighbourhoods**. Six statistical descriptors can be used: *mean, mode, median, standard deviation, range* and *skewness*. They are designed to inform on the variations of backscattered **intensity** and **multi-echo LiDAR features**: *return number, number of returns* and their ratio called *echo ratio*. Combined with the six statistical

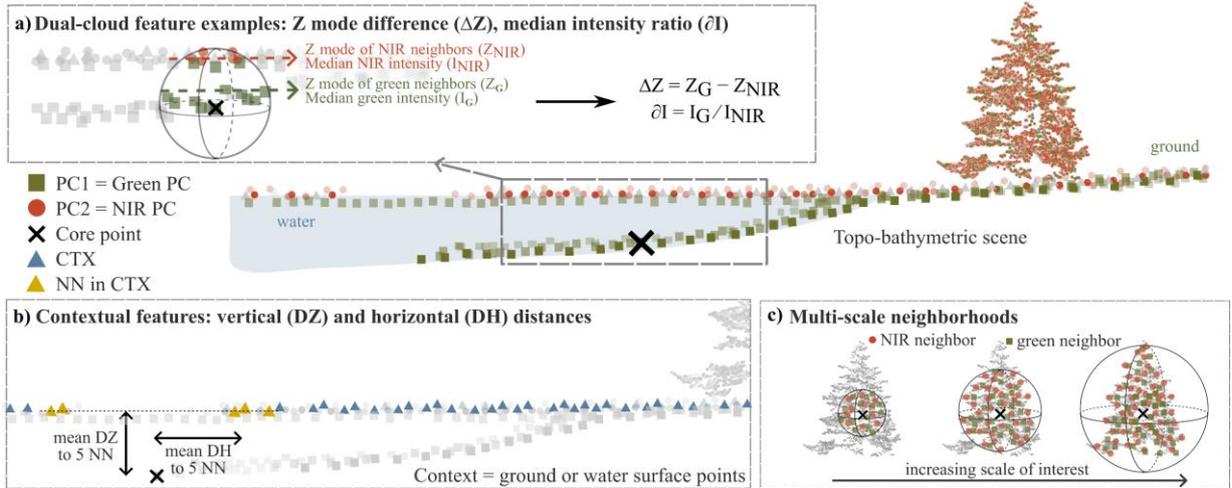

*Figure 2: Illustration of the main characteristics of 3DMASC: a) examples of new dual-cloud features providing a better description of the differences between clouds, b) the generalized contextual attributes placing each point in its spatial setting and c) the multi-scale spherical neighbourhoods used to describe the many aspects of 3D objects*

descriptors, these four point-based features result in 24 neighbourhood-based features at a given scale. To our knowledge, these types of rich multi-scale statistics were never used before for raw 3D PC classification. Similar features can be built from the three components of the **RGB** colour information, and we evaluate the benefits of this information for classification at a later stage.

### 3.1.3 Dual cloud features

Dual cloud features describe the geometrical, spectral, height statistics or multi-echo characteristics differences between the neighbourhood of the core point in PC1 and PC2. Spectral ratios have been introduced in the context of multi-spectral lidar classification (Chen et al., 2017; Morsy et al., 2017b; Wichmann et al., 2015), but geometrical, height statistics and multi-echo characteristics are new contributions. We designed them to leverage the bi-spectral information and improve the descriptions of scenes characterized by a different 3D aspect in PC1 and PC2. In TB lidar datasets, the NIR and green PCs are most significantly distinct above water and vegetation (Figure 1), but they can also be slightly different over other surfaces. This is due to the different surface optical characteristics and the NIR and green laser emitters that can have different angles of incidence or aperture. These may cause differences in the returned signal intensity and the 3D position of the points. The definition of these features assumes that both PCs are correctly registered and that the alignment error is as low as possible for geometric differences to be related to objects' characteristics and not registration errors.

Dual-cloud features result from **mathematical operations** between single cloud features of the same core point's neighbourhood in PC1 and PC2. They can be feature *differences, additions, multiplications*, or *divisions*. Here, we have used *differences* to measure dissimilarity, in particular for elevation, geometry and multi-echo features, and *divisions* to normalize one feature by another, typically for intensity. Figure 2a illustrates two examples of dual cloud features: the *mode difference of elevation* that is expected to be close to zero on the ground but different over water; and the *median intensity ratio* between the green and NIR channels that is expected to be distinct over different grounds. A selection of dominant features is presented, illustrated and explained in the Results section. Dual cloud features also encompass a distance computation (vertical or horizontal) between the core points PC and another PC (PC1, PC2, or CTX), using kNN.

### 3.2 Random Forest Classification

3DMASC uses a Random Forest (RF) algorithm (Breiman, 2001) to perform PC classification, i.e., predict a label y ∈ $\{1,2,…,c\}$ for each point $P_k$ of the input PC, using the predictor vector F $\{F_{kij} \mid k=1,…,n; i=1,…,f; j=1,…,s\}$, where *f*=number of features, *s*=number of scales and *n*=number of core points and each $F_{kij} \in \mathbb{R}$. For instance, the label can represent the type of object sampled by P.

Here, the feature importance is the product between the probability of reaching a node (i.e., the proportion of samples that get that node) and the Gini impurity decrease of that node. Feature importance is normalized to sum up to 1. A higher value symbolizes a more significant influence of the feature on the prediction.

RF does not handle Not-a-Number (NaN) values, which may be present with our features (depending on the scale, NaN values can occur if the spherical neighbourhoods do not contain the minimal number of points required to compute the covariance matrix, for example). This requires specific pre-processing of the predictor vector. Indeed, replacing NaNs with a fixed value may imply irrelevant representations of the local sub-clouds and thus incorporate bias in the classifier training. To tackle this issue, 3DMASC relies on the RF implementation of the cross-platform library OpenCV (Bradski, 2000), which incorporates surrogate splits to handle missing measurements. We use base settings and forests populated with 150 decision trees, having a maximum node depth of 25, as advised by Oshiro et al., 2012. We also compared it with the RF implementation in the Python library Scikit-learn (Pedregosa et al., 2011) and found similar results.

To further improve the classifiers' robustness, we exploit the prediction probability output by RF and use it as a classification confidence indicator, as seen in (Brodu and Lague, 2012) and (Letard et al., 2022). The prediction probability corresponds to the proportion of forest trees that voted for the class assigned to the point. It ranges between 0 and 1.

### 3.3 Features and scale pre-selection to control the size of the predictor vector

We propose a feature selection routine (Dash and Liu, 1997) to improve the explainability and efficiency – through the number of predictors – of the trained algorithm, as there can be almost 90 features per scale in TB environments.

Although information redundancy supposedly does not impact RF performances, it disrupts the explainability of the feature importance values since if two features bring similar information, their relative importance will be underrepresented. Thus, we keep only a set of uncorrelated features by using a bivariate feature selection (Dash and Liu, 1997; Guyon and Elisseeff, 2003), incorporating an assessment of the features' Information Gain (IG) (Dash and Liu, 1997) and the Pearson linear correlation coefficient of attribute pairs. The correlation threshold and the scale at which each feature is evaluated are user-defined and determined after an empirical investigation.

The same bivariate procedure allows the selection of scales. However, we also decided to promote small scales to limit the computation cost of the classifier. The selection process relies on a majority voting procedure. Since it is impossible to consider a scale independently from its application to a feature, we retain the scales that are the most often selected when they are evaluated for each feature independently.

Considering the variety of features included in 3DMASC, removing correlated features and scales does not provide a significantly smaller set of features. Typically, in our experiments on two datasets developed in the next sections, around 40 features per scale of interest remain after correlation-based pruning. The classifier obtained may thus not be easier to explain, and the application steps may be unnecessarily computationally heavy.

To further reduce the dimension of the predictor vector, we considered a feature ranking depending on the IG. However, defining a fixed number of features and scales is highly task- and site-dependent, and filter-based selection would not consider internal synergies between features. Consequently, we use an embedded backward feature selection, relying on the RF feature importance, as detailed in (Aggarwal, 2014; Dash and Liu, 1997). This selection is performed on the uncorrelated set previously obtained. The optimized predictor vector is then identified through automatic out-of-bag score (OOB) monitoring. OOB corresponds to the performance score obtained by the RF algorithm on a sample that is not part of the subsample of data used to build the trees (Breiman, 2001). It is used to estimate the generalization ability of the model, i.e. the performance of the classifier on new, unseen data. In practice, we tested the performance of each model build during optimization on our test dataset and indeed observed that prediction accuracies on the test data varied similarly to OOB (see Supplementary Materials, Figure 2). We thus use a sliding window to monitor the variations of OOB over a given number of iterations (in our experiments, ten) and keep the last best iteration before OOB starts to drop (i.e. varies more than a user-defined threshold). In the rest of the paper, we will refer to this step as classifier optimization, which, since the OOB score is obtained using the training data, is performed completely independently from the test data.

### 3.4 Framework implementation

Figure 3 summarises the global framework introduced in this work and illustrates how the different steps explained follow each other when processing a PC. As detailed in Appendix A, the Cloudcompare 3DMASC plugin can be used at two levels of complexity: for beginners, a complete graphical user interface (GUI) exists from feature computation to classifier training and class inference; for expert users, 3DMASC can be called through command line solely for fast feature computation with its parallelized C++ implementation, and the results subsequently used in any other environment such as python. Features and scales preselection and classifier optimization described in section 3.3 follow this latter approach and operate through a complementary Python script. To avoid feature preselection and classifier optimization for non-specialist users, this work aims to identify a minimal set of features and scales that can systematically be used for TB lidar classification.

## 4. DATASETS AND EXPERIMENT PROTOCOL

### 4.1 Experimental datasets and classes

To illustrate the use of 3DMASC for bispectral lidar data classification, we selected two topo-bathymetric lidar datasets, representing one coastal and one fluvial environment, respectively (Figure 1). These two datasets only differ in the type of environment they model. They were both surveyed with a Teledyne-Optech Titan airborne lidar with two wavelengths, 532 nm and 1064 nm (Lague and Feldmann, 2020). The green laser points with a forward pitch of 7° necessary to avoid strong surface reflection on water and has a beam divergence of 0.7 mrad. The NIR laser has no forward pitch and a beam divergence of 0.3 mrad. Consequently, the incidence angle, surface sampling and laser spot size are never the same for the two lasers at a given scene location. The sensor produces high-density PCs, typically 36 pts/m² on land – when combining both wavelengths – and 18 pts/m² under water in a single pass (Lague and Feldmann, 2020). More details about the sensor and the acquisition conditions – typical aircraft altitude, speed, overlap between flight lines and preprocessing – are available in (Lague and Feldmann, 2020). The mean vertical offset between the two channels measured on flat horizontal surfaces is typically less than 1 cm. The precision evaluated as the standard deviation of point elevation measured on flat horizontal surfaces is around 5 cm on topography and 10 cm on submerged surfaces. The first site lies on the French coast of the Channel, in Britanny, near Fréhel; the second is a portion of the Ain River in South-Eastern France near its confluence with the Rhône River. The surveys were conducted in May 2021 and September 2016, respectively. Figure 1 features the two scenes. They both contain natural and anthropic land covers and include a part of the bathymetric environment: in the first case, shallow sea water with green laser extinction at 10.5 m; in the second case, a river with green laser extinction at 3.5 m. The flights combined lidar surveys and simultaneous RGB imagery acquisitions with the control camera, which produced orthoimages with ground sampling distances of 25 cm and a registration error of about 20 cm. As RGB imagery acquisition was not the main objective of the surveys, pronounced shadows exist, particularly on the Ain survey, as it happened late in the afternoon.

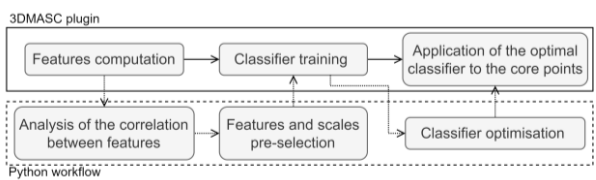

Figure 3: Illustration of the 3D Multi-Attributes, Multi-Scale, Multi-Cloud (3DMASC) classification workflow.

| Basic classes (primary classification) | Detailed classes (advanced classification) | |
|---|---|---|
| Both | Ain (river) | Fréhel (coast) |
| Ground | Bare ground | Sand |
| | | Pebble/cobble |
| | | Rock |
| | Artificial ground | Artificial ground |
| | Vegetated ground | Vegetated ground |
| Vegetation | Intermediate | Vegetation |
| | High | |
| Artificial elements | Buildings | Buildings |
| | Power lines | Power lines |
| | Vehicles | Vehicles |
| Seabed/ Riverbed | Riverbed | Sandy seabed |
| | | Rocky seabed |
| | Swimming Pools | / |
| Water | Water column | Water column |
| | | Surf zone |

Table 1: List of classes defined for the experiments.

### 4.2 Classes definition and 3D annotation

We evaluate the performances of 3DMASC on two levels of detail: a *primary* classification of 5 land covers – strictly identical for both areas – and *advanced* labelling of 11 and 13 types of objects on the Ain and Fréhel datasets, respectively. We chose the classes depending on the diversity of land and sea covers we could observe in each area. Table 1 contains all the categories that we use for the primary and advanced classifications. *Artificial ground* includes roads and surfaces covered with concrete or tar (parking lots, dykes). *Vegetated ground* is grass or other low vegetation, such as low-growing heather in moors. In the Ain survey, *intermediate vegetation* is defined as bushes or shrubs with a different aspect than high trees and a smaller growing height. We did not use a classical classification based on a strict height threshold, as usually made in vegetation mapping applications (Letard et al., 2022). Our objective was to avoid the traditional misclassification of low branches attached to high trees as shrubs while they are points belonging to high vegetation. The definition of *intermediate vegetation* and *high vegetation,* therefore, balances the 3D aspect and height above ground. Compared to other classes that can be objectively defined, our separation between intermediate and high vegetation is somewhat subjective. The lack of various types of vegetation in the Frehel datasets prevented us from refining the vegetation class. We annotated portions of data manually using visual interpretation of the PCs and the RGB imagery acquired simultaneously using Cloudcompare (Girardeau-Montaut, 2022), including new specific developments for quick labelling of 3D point clouds. Four training and test datasets – one for each level of detail of each scene – were created, all labelled and balanced, for the classification experiments. They all contain 2000 points of each label. To eliminate potential spatial bias due to the use of multi-scale spherical neighbourhoods, we forced each training and test point of the same label to be at least 20 m away, considering we used spheres with diameters up to 15 m. Figure 4 illustrates the resulting sets of points labelled for training and testing. The annotated datasets are available along with the plugin's source codes and the scripts used to perform further analysis at the following link: https://github.com/p-leroy/lidar_platform. These datasets contain the NIR and green high-density surveys of both areas, a context PC representing a raw ground detection of each site at 2 m spacing, and the annotated training and test points, with two levels of detail, one with five broad classes, another with 11 or 13 detailed classes depending on the area.

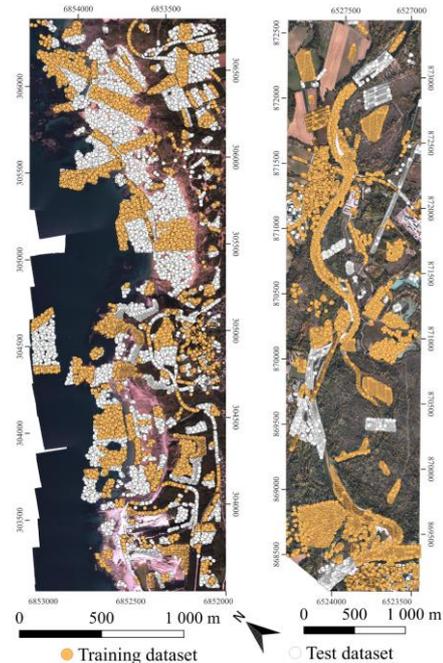

Figure 4: Location of the train/test data in the surroundings of Fréhel (left) and of the Ain (right) (RGF93).

### 4.3 Evaluation metrics

We use the **Overall Accuracy** (OA) to quantify the correct proportion of global predictions. The **precision** estimates, for each label, the correct proportion of positive predictions. The **recall** value evaluates the part of true positives identified correctly (Sokolova et al., 2006). Precision thus tends to outline over-estimation of some classes, while recall highlights under-estimation. Often, the goal is to balance precision and recall so there is no clear tendency of over- or under-estimation of the classes considered. In a context where under- or over-estimation of a given class is not targeted, the smaller the difference between both metrics, the better the result. The **F-score** combines the information provided by precision and recall through their harmonic mean (Sokolova et al., 2006).

In addition to performance quantification, we wish to address the explainability of the method. According to Roscher et al., 2020, explainable machine learning in the natural sciences should incorporate transparency, interpretability, and explainability. We address each of these elements by detailing the method, exploring feature and scale importance and selection mechanisms, and analysing explicative elements of the decision rules behind the results obtained and their reproducibility among different environments. The class-wise performances are explainable with the approach of Lundberg et al. (2017) by computing the **Shapley values** (Shapley, 1953). These range between 0 and 1 and quantify, for each point, the influence of each feature on the label prediction based on game theory concepts. We performed this analysis using the SHAP Python library (Lundberg et al., 2017). Using these values as a complement to the variable importance measurement and a low number of predictors in the optimized models allows us to have a more robust explanation, less dependent on the randomness of descriptors and samples selection at each node of the trees.

## 5. RESULTS

This section first presents the overall classification results obtained in the fluvial and coastal environments and the impact of feature preselection and optimization. We then present the class-wise results, dominant scales, and features that emerge from the experiment. Finally, we explore the benefits of using RGB information, contextual data and classification capabilities when using only green lidar data. All results presented are obtained on a test dataset strictly different from the training dataset.

### 5.1 Overall classification results depending on the number of predictors

We use three different terminologies: Full predictor Set (FS) classification implies the computation of all features at all selected scales; Optimized Classification (OC) does not consider all features at all scales; Single Scale Classification uses an identical single scale for all feature calculations. The starting set of features contains 88 features, which include all possible features of 3DMASC computed on PC1, PC2 and their difference or ratio between both PCs (see Appendix B). They are calculated at 29 different scales from 1 m to 15 m with a 0.5 m increment and for kNNs with k in {1;2;3;4;5;10}. The complete predictor vector has 2011 columns (4 point-based features, 23 features computed at 29 scales for three spherical neighbourhoods – green, NIR, both – and six kNN-based features).

To determine the scale to use for feature evaluation – i.e. IG assessment – we analyzed the OA obtained when selecting features based on their IG at scales varying from 1 m to 12 m. This first analysis shows that features computed at 2 m allow the best selection for OA (see Supplementary Material). Similarly, testing for the optimal correlation threshold results in a value of 0.85 (see Supplementary Material). A scale of 4 m would have been valid, too, for the Ain dataset, but since there was no difference between using 2 m and 4 m for this area, we picked 2 m to ensure the comparability of the results between the two zones.

#### 5.1.1 Impact of correlation and feature pre-selection

The same feature computed at two scales separated with a small gap is expected to produce redundant information. Here, we explore if all types of features exhibit similar levels of correlation with scale. Figure 5 presents the mean absolute Pearson coefficient between features computed at scales separated by 1 m or 3 m for different types of single cloud or dual cloud attributes: geometrical, echo-based, and intensity-based. The general tendency is for absolute linear correlation to increase with scale and to saturate or increase only slightly above a threshold scale of about 4 to 6 m. The maximum linear correlation level depends on the type of feature and environment. Dual-cloud geometric features are less correlated than single-cloud features. Intensity-based features exhibit high linear correlation levels, suggesting a potentially substantial redundancy across scales. The comparison between the dual cloud geometric features at Fréhel for steps of 1 m and 3 m shows that the larger the step between scales, the lower the linear correlation. As expected intuitively, the step between scales should thus tend to increase with scale, particularly above 6 m, to limit information redundancy.

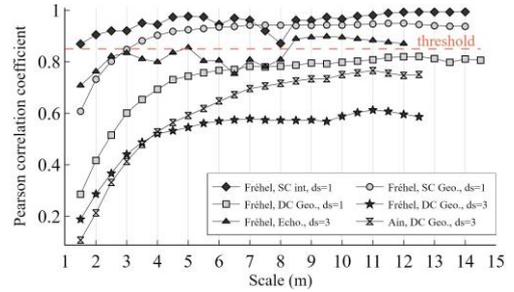

Figure 5: Linear correlation between features computed at scales separated by ds=1 m or ds=3 m for different examples of features. SC = Single Cloud; DC = Dual Cloud. The threshold at 0.85 emerged from an empirical analysis.

These results indicate that given the high linear correlation of certain features, especially above 4-6 m, there is hardly a need for many individual scales above this scale, particularly for single cloud intensity and geometric features. Consequently, we enforce the maximum number of scales kept in the preselection phase to be 10, compared to the initial 29. Finally, Figure 5 demonstrates that intra-feature scale correlation is site-dependent and that no clear principles rule correlation dynamics. Consequently, it is impossible to select scales for features based on their linear correlation without first computing them.

After feature preselection accounting for absolute linear correlation, different features are eliminated depending on the site. Overall, there were fewer correlated features on the Ain site and more correlation when using a higher number of classes (and therefore feature samples). The number of features passing the selection step ranges between 36 (Fréhel, primary classification) and 44 (Ain, primary classifier). Height-derived and dimensionality-based attributes were the most pruned types of features during correlation filtering. NIR and green roughness and return numbers are strongly correlated in both areas. Measures of echo ratio were too correlated in the Ain but not in Fréhel, which reflects the differences between riverine and coastal TB surveys.

#### 5.1.2 Impact of predictors number and optimization: from full predictor set to optimized classifiers

We explore the influence of the number of features and scales used on the OA and present the results in Figure 6. The results confirm the conclusions of (Brodu and Lague, 2012; Thomas et al., 2018) on the superiority of multi-scale algorithms compared to single-scale classifiers. This analysis also illustrates the decreasing benefit of increasing the number of features and scales past 20 features and 6 scales, even using uncorrelated entities only. Figure 6 highlights that adding features increases OA more than adding scales. For instance, adding a second scale to a single feature classifier systematically results in an OA surge, while harvesting 10 features at the same two scales produces more accurate results than relying on two features computed at the same 10 scales.

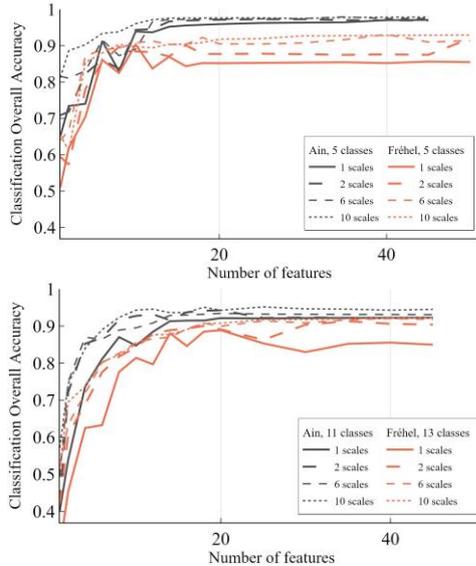

Figure 6: Classification accuracy depending on the number of features computed at different numbers of scales.

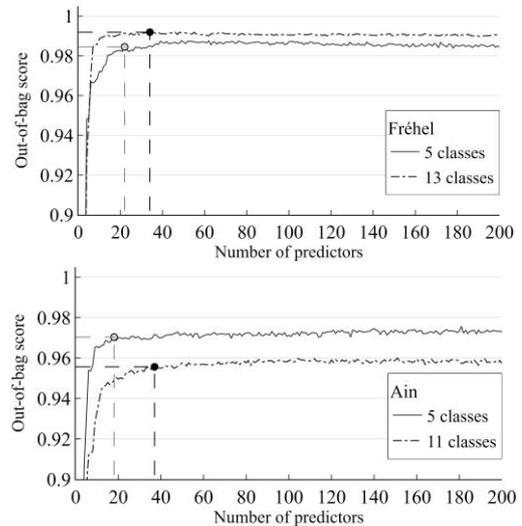

Figure 7: Out-of-bag score depending on the number of predictors used during classifier optimization.

Due to the majority voting used for scale selection, the scale used for single-scale classification varies, explaining the accuracy variations (see Figure 6, single-scale curve) and showing the dependence between the features' relevance and their computation scales.

Since the accuracies presented are the results of applying the trained algorithms to data unseen during training, these results also showcase the stability of RF relative to overfitting and generalization. Even when training the model with hundreds of predictors, OAs remain stable (between 92% and 98%, depending on the use case) when classifying the distinct test points (see Figure 6). Furthermore, 3DMASC's features succeed at characterizing the nature of the objects lying behind the points, as accuracies converge towards values ranging between 92% (Fréhel, advanced) and 98% (Ain, primary). It is, however, delicate to determine the ideal number of features and scales to retain. The optimization procedure provides more information on the required number of predictors to achieve high-accuracy identification of the different classes. Figure 7 presents the OOB evolution when reducing the predictor set iteratively. With the automatic monitoring of the OOB's significant variations, a set of parameters is chosen, providing optimized classifiers for the four experiments and corresponding training datasets. Table 2 gathers the main characteristics of the optimized classifiers.

The results in Table 2 confirm what we observed in Figure 6: a small number of features and scales produces highly accurate classifications. The most complex classifier incorporates 37 predictors, including 16 features and nine scales. Table 2 also outlines that more predictors are needed to correctly identify more labels: advanced classifications require 19 and 12 more predictors than primary for the Ain and Fréhel areas, respectively. The optimized models obtain accuracies ranging between 90.7% and 97.6% and harvest more features than scales, confirming the superior efficiency of feature diversity over scale abundance. Overall, the maximal difference in OA between full set and optimized classifiers is 1.2%. Models are highly simplified: on average, the optimization reduces the predictor vector's dimension by 93%. However, the fully iterative procedure is necessary to determine the number of predictors to use and limit the loss of OA. For example, when using the 18 highest-ranked predictors at the first RF classification of the Ain, the OA is 94%, almost a 4% difference.

| Classifier | Ain (5 cl) | Ain (11 cl) | Fréhel (5 cl) | Fréhel (13 cl) |
|---|---|---|---|---|
| FS OA | 97.9% | 94.6% | 92.8% | 91,9% |
| FS nb of pred. | 371 | 352 | 315 | 330 |
| OC OA | 97.6% | 94% | 91.6% | 90.7% |
| OC Nb of pred. | 18 | 37 | 22 | 34 |
| Features | 12 | 16 | 12 | 19 |
| Scales | 5 | 9 | 6 | 6 |
| Mean confidence | 0.93 | 0.9 | 0.89 | 0.83 |

Table 2: Characteristics of the four models. FS = full predictor set, OC = optimized classification. Nb of pred. refers to the number of predictors used.

### 5.2 Class-wise results with optimized classifier

#### 5.2.1 Class-wise metrics

Figure 8 illustrates the application of the optimized classifiers for the *advanced* classification. The land-water transition is well identified, and the main elements, such as ground and above-ground features, are separated. Figure 9 sums up the class-wise results obtained for each experiment. The main classes of the Ain site obtain F-scores higher than 85%. In the coastal area, they are distinguished with F-scores over 88%.

The difficulty imposed by the distinction of objects with similar geometries does not impact the performances severely.

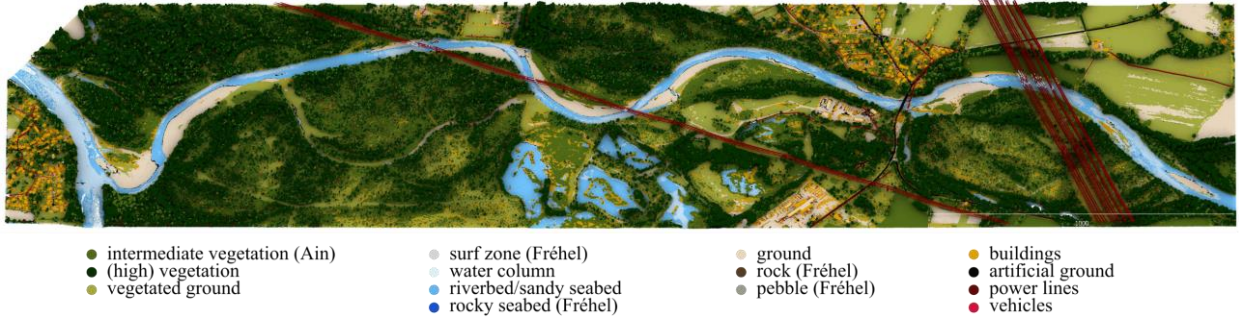

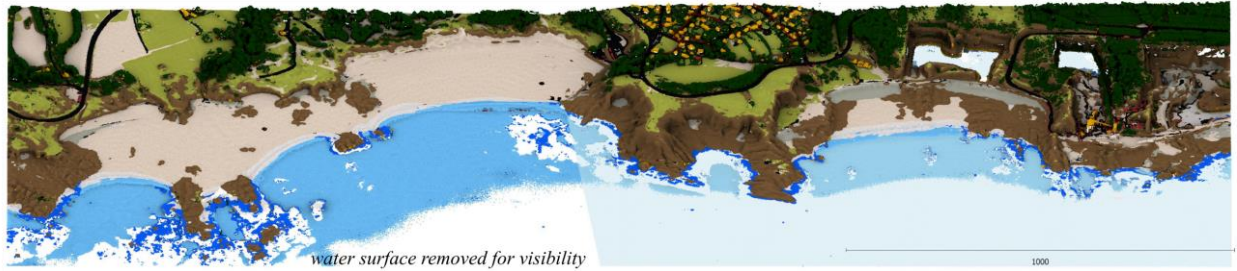

Figure 8: Classified point clouds of both areas using the optimized classifiers, with Fréhel on top and the Ain under.

All F1-scores are above 83%, and average confidences are over 70% and 80% in primary and advanced cases, respectively, except for *vehicles* in the advanced Fréhel experiment. These observations suggest efficient construction of the classifiers, as correct predictions obtain the vote of most of the decision trees.

The identification of *water column* is highly accurate (99%) in both advanced classifications, but there is more confusion in the primary experiments, where the broader classes may be harder to define. The *surf zone* is also challenging to distinguish from *ground* or *rocky seabed* in some areas of Fréhel. Some classes show gaps between precision and recall, reflecting the over-detection of *buildings* in rocky areas or of *intermediate vegetation* in the Ain (see Figure 8).

### 5.2.2 Dominant scales analysis

The optimized predictor vectors indicate that some features are particularly informative at specific scales, and conversely, some scales are essential for given features only. The optimization phase alters the systematic multi-scale character of the classification since the number of predictors in the optimized models is smaller than the product between the number of scales and the number of features. For example, the advanced classification of the Ain has an optimized predictor vector exploiting 16 features at nine scales, yet its total size is 37. In contrast, if the optimized classification systematically used all available scales for a feature, it would be 144.

| Classifier | Optimized set of scales |
|---|---|
| Ain, primary | 1.5 m, 4 m, 5.5 m, 14 m, 10NN |
| Ain, advanced | 1.5m, 2 m, 2.5 m, 3 m, 3.5m, 4 m, 5.5 m, 7.5 m, 5NN |
| Fréhel, primary | 3 m, 3.5 m, 5.5 m, 6 m, 14.5 m, 1NN |
| Fréhel, advanced | 3.5 m, 4.5 m, 6.5 m, 7 m, 14.5 m, 1NN |

Table 3: remaining scales in the four optimized multi-scales classifiers. kNN indicates k nearest neighbours.

Table 3 sums up the specific scales retained for each experiment. It shows that finer scales are necessary to describe the Ain site: the minimal scale selected is 1.5 m, whereas it is double for Fréhel. All classifiers follow a similar pattern: they exploit small to medium scales up to about 6 m and a much larger scale of about 14 m without transitioning via a medium value. The advanced models both reuse similar scales to their primary equivalents but incorporate new ones in between, reducing the typical sampling step of object sizes. However, the 11-label classifier of the Ain is the only one discarding the 14 m scale, thus exploiting only small to medium diameters.

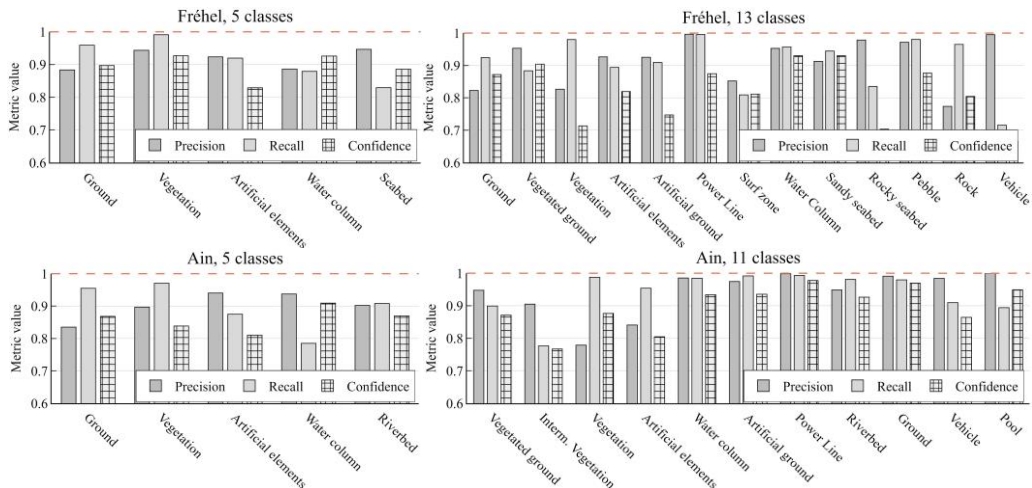

Figure 9: Precision, recall, and prediction confidence per class for the four classifiers after optimization.

| AIN | | FREHEL | |
|---|---|---|---|
| RGB Image | | RGB Image | |
| Z difference with NIR kNN | | Z difference with NIR kNN | |
| Third eigenvalue (green) | | Sphericity (green) | |
| Difference of Z modes | | Ratio of median intensities | |
| Standard deviation of Z (green) | | Skewness of intensity (NIR) | |
| Standard deviation of Z (NIR) | | Mode of intensity (green) | |
| Difference of roughness | | Mode of intensity (NIR) | |
| Mean intensity (green) | | Mean intensity (green) | |
| Mean EchoRatio (green) | | Mean EchoRatio (green) | |
| Mean Nb of Returns (green) | | Mean EchoRatio (NIR) | |
| Mean Nb of Returns (NIR) | | Mean Nb of Returns (green) | |
| | | Mean Nb of Returns (NIR) | |
| | | Mean Return Nb (green) | |

Table 4: Optimized set of features for both sites. For dual cloud features corresponding to a difference, blue values are negative, white is zero, and red is positive. For other features which are strictly positive, blue indicates lower values, green intermediate values, and red higher values.

To better identify the contribution of specific scales to various classes in the two environments, Figure 10 shows the Shapley analysis for the standard classification. Dominant scales are different between the Ain area and Fréhel. Water column and seabed/riverbed are dominated by features computed with around 3 m, 6 m and 14.5 m diameter in Fréhel, whereas 6 m and 10NN features are more useful in the Ain. Similarly, *artificial elements* and trees do not exploit the same sphere sizes over the two sites. The scales also adapt to each label. For example, *artificial elements* – buildings, vehicles, and power lines – rely less on kNN features than riverbed in the Ain or *ground* in Frehel.

We can also identify two groups of classes having similar scale contribution patterns. The first includes *water column* and *seabed/riverbed*, and the second includes *ground* and *artificial elements* in Fréhel, while it is composed of *ground* and *vegetation* in the other area. In Fréhel, *vegetation* follows similar trends as the bathymetric classes.

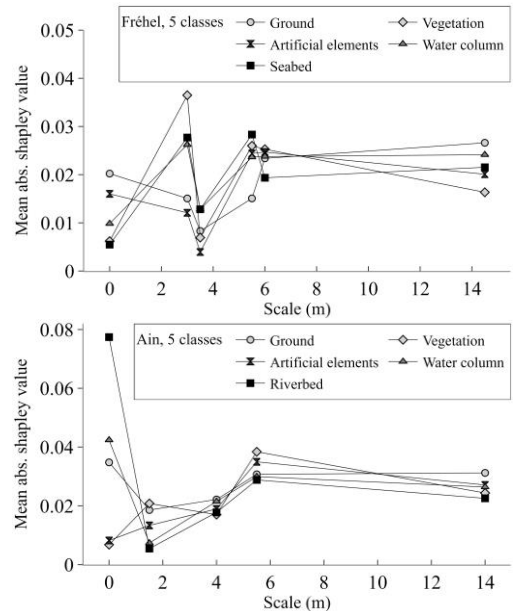

Figure 10: Mean absolute Shapley value of each scale of the optimized predictor vector depending on the class considered (0 m scales represent features computed with a kNN search).

### 5.2.3 Dominant features analysis

To simplify, we only review the dominant features of the primary classifications in this section. Several features stood out from the rest and passed both the selection and optimization phases. They theoretically contain the essential information to distinguish the defined classes. Table 4 introduces and illustrates each of them.

The optimized sets of predictors obtained, presented in Figure 11 and Table 4, seem to be tailored to each site. The Shapley analysis in Figure 11 corroborates this observation. **Only five features common to both sites are identifiable**: *vertical*

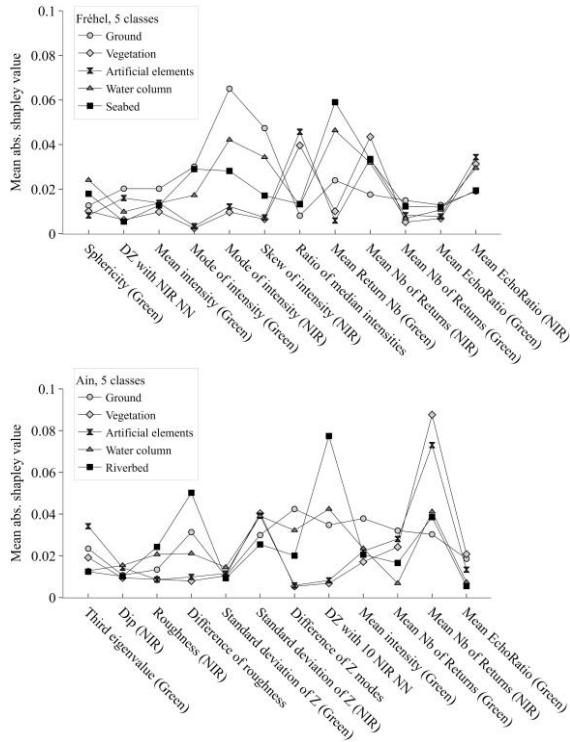

Figure 11: Mean absolute Shapley value obtained by each feature of the optimized predictor vector depending on the class considered

*distance of green points to their NIR neighbours (kNN), mean green intensity, mean echo ratio in green neighbourhood, mean number of returns in the green neighbourhood, and mean number of returns in NIR neighbourhood*. Two **groups of labels have similar feature contribution patterns**. *Ground*, *seabed/riverbed,* and *water column* on one side and *vegetation* and *artificial elements* on the other. Multi-echo features, NIR intensity, and dual-cloud features mainly identify the first group. The second relies primarily on dual cloud features – median intensity differences – and NIR multi-echo attributes.

In both cases, **the TB aspect of the datasets is fully exploited**: in both areas, four NIR PC features and five to six green PC attributes are involved in the optimized sets. NIR PC-derived features are more contributive to topographic objects, while both PCs are equally crucial for ground/seabed/water column distinction. The experiments on Fréhel also draw more on NIR intensity-derived parameters than the models to process the Ain, in which only one green spectral parameter is involved with low relative importance (Figure 11). The class-wise feature importance analysis also shows that **features do not have the same descriptive power in both NIR and green domains**. The number of returns of the NIR echoes is more informative on the nature of the surface than its green equivalent.

Both results show a **predominance of newly introduced 3DMASC features over classical features** used in other studies (Chehata et al., 2009; Hackel et al., 2016; Thomas et al., 2018; M. Weinmann et al., 2015). 8 out of 12 for the Ain site and 10 out of 12 for Fréhel are attributes we propose with 3DMASC: means, modes, or skewness values of PC characteristics, as well as dual-cloud features. Geometrical and dimensionality-based features are scarce: only NIR PC roughness, NIR PC dip, green PC sphericity, and green PC third eigenvalue pass the optimization phase. The mean green intensity is the only other example of classical feature observable (see Table 4). Intensity-based features constitute nearly half of the predictors of the Fréhel optimized classification but are few in the Ain model. The other half of the Fréhel predictor set is dominated by multi-echo features of both wavelengths that evict height-based features and geometrical features. In the Ain, they appear through the differences in elevation modes. Other dual cloud features stand out: *vertical distances between green and NIR points*, *elevation mode differences*, and *median intensity differences between the two PCs* (Table 4).

Figure 11 also reveals that the **new 3DMASC features outperform usually dominant characteristics** like intensity. These features are the new features we introduced: means, modes, skewness or standard deviation of existing attributes as well as dual cloud features. The difference in elevation modes between NIR and green PCs is more relevant for identifying vegetation than intensity, in particular in the Ain. Similarly, the roughness difference between PCs dominates single cloud NIR roughness, even for ground and over-ground object separation. The ratio of median NIR and green intensities is beneficial for outlining vegetation and artificial elements. Dual cloud features are present in both OC classifiers, illustrating how they complement separate single cloud attributes. Multi-echo features also contribute significantly to the predictions. The mean number of returns is handy for characterizing *vegetation* and *artificial elements*.

**5.3 Results using other predictors**

In this section, we test 3DMASC in different settings: using a context PC, RGB information, and simulating the unavailability of the NIR wavelength. All results are summed up in Figure 12. They are obtained by running the complete framework on initial predictor vectors, including contextual, RGB-derived, or green features only. The presence or absence in the optimized predictor set of each tested attribute is thus already an indication of their informative character.

The contextual features used were vertical distances to a PC containing only ground or water surface points, for different scales (1 m, 3 m, 5 m, and 10 NN). These predictors allowed to use smaller scales (see Supplementary Materials) and improved the prediction confidence and quality of all classes in the Ain area. In Fréhel, they improved the accuracy of *seabed*, but tend to penalize *water column* and *vegetation*.

The reflectance in the blue domain is the only RGB derived attribute that passed optimization. Its mode is used in two models: Fréhel primary, and Fréhel advanced. This shows that RGB features are not crucial to detect the classes of the Ain but may serve to differentiate coastal land and sea covers. They also seem to penalize our classifier optimization framework when they are used but do not appear in the best models, as the losses in F-scores on the advanced classification of the Ain reveal. This shows that RGB parameters may evict more valuable features and reduce the classifier's abilities on unseen data. However, in the primary case in this area, F1 scores are higher when the optimization is performed on this extended feature set: when including the RGB attributes, the OOB significantly drop when the number of predictors is below 40, causing the presence of additional intensity-based 3DMASC attributes in the optimized classifier, which seems to increase F1 scores of several classes.

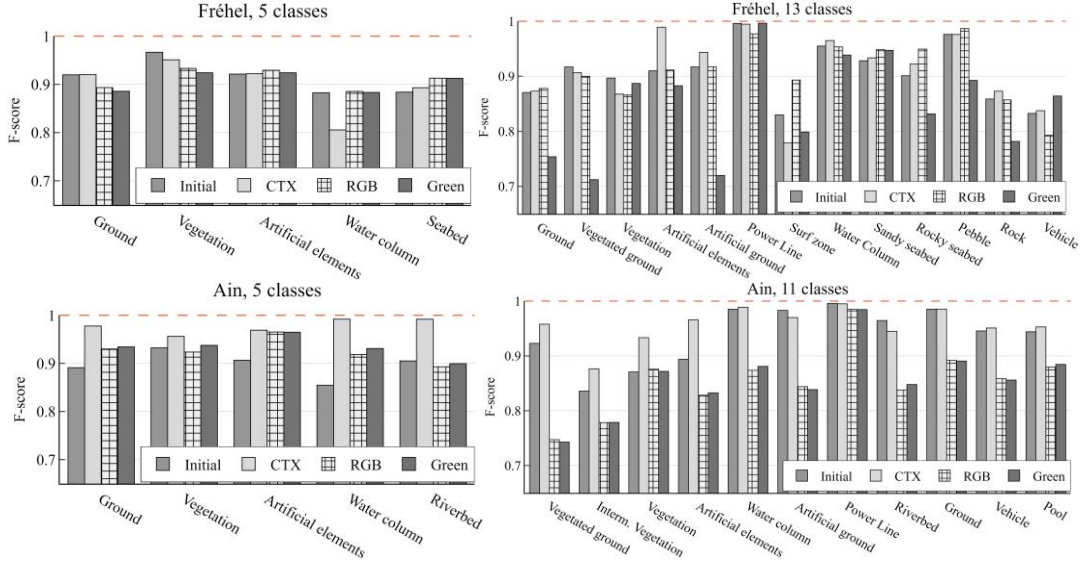

Figure 12: F-score obtained for each class depending on the experiment. Initial = optimized classifier obtained with the initial set of predictors. CTX = optimized classifier obtained when adding contextual features. RGB = optimized classifier obtained with RGB features added. Green = optimized classifier obtained using only green features.

When using green laser data only, OAs range between 85% (Fréhel, advanced) and 94% (Ain, primary). Multi-echo features and intensity-derived attributes dominate predictor vectors. In Fréhel, dip and standard deviation of intensity are the only new features selected. In the Ain, point-based echo ratio, mean return number, mode, and standard deviation of intensity appear. Overall, more scales are used per feature, and seven and eleven features are selected for Fréhel and the Ain, respectively. In the 5 class experiment of the Ain, a performance decline is only observed for *riverbed*. Although its F-score drops by 2%, it remains at 90%, showing that a single bathymetric PC already provides highly accurate detections of the water column and the riverbed. In Fréhel, the classification of *seabed* is even improved when excluding NIR data. In both settings, the distinction of topographic classes is less accurate when discarding NIR information.

## 6. DISCUSSION

Starting with a set of 88 features computed at 29 scales, we obtained optimized, compact classifiers ranging from 18 to 37 predictors – scales and attributes – resulting in good to excellent classification for up to 13 classes. In this section, we discuss these results with respect to existing work on PC classification.

### 6.1 Classifier optimization and number of predictors used

Through 3DMASC, we obtain classifications of TB scenes with OAs over 90%, using light classifiers that harvest a maximum of 37 predictors (Table 2), some of them using as little as 18 predictors (Ain, five classes). Average prediction confidence is high and accompanied by high accuracy, a synonym of effective classifier training. Low confidence values can be linked with classification errors and used to filter out misclassified points. Table 5 shows the results of applying a confidence threshold below which points are removed. It illustrates that there is necessarily a balance to find between result quality and spatial resolution of the classified PC, as aiming at fewer classification errors means accepting to reduce the local density of the data.

| CLASSIFIER | Confidence threshold | OA | Remaining points (%) |
|---|---|---|---|
| **AIN advanced** | 0.5 | 95% | 96% |
| | 0.6 | 97% | 92% |
| | 0.7 | 98% | 87% |
| | 0.8 | 98% | 80% |
| **FREHEL advanced** | 0.5 | 94% | 92% |
| | 0.6 | 96% | 84% |
| | 0.7 | 97% | 76% |
| | 0.8 | 97% | 67% |

Table 5: Overall accuracy depending on the confidence threshold applied to filter the predictions.

The optimization step seems to balance computational efficiency and high-quality classifications. The low number of predictors makes the models applicable to large datasets, easily explainable with Shapley values and thus accessible to non-specialist users. These characteristics allow 3DMASC to be an interesting alternative to current state-of-the-art methods that are 3D deep neural networks. As mentioned in section 2.1, discussing the performances of deep neural networks is out of the scope of this study, but they have proven to output significantly good results on 3D semantic segmentation applications (Y. Guo et al., 2021). However, to our knowledge, no available and accessible deep learning framework exists for multiple 3D PC classification or for airborne lidar data in similar natural settings that environmental scientists could easily reuse without a significant background in deep learning. Indeed, the developments still primarily focus on urban areas (Huang et al., 2021; Lin et al., 2021; Mao et al., 2022a; Schmohl and Sörgel, 2019; Wen et al., 2021; Yang et al., 2018; Zeng et al., 2023; Zhang et al., 2022; Zhao et al., 2018), where the variety of scales and geometry is quite dissimilar to what we observe in natural, TB settings. Neural network hyperparameters are also harder to optimize without expert knowledge and require more complex and intensive training, and thus, computing power and are thus less easy to master for thematic users. In this context, we addressed the need for an approach allowing for high-accuracy multi-class classification while relying on less complex computation and staying accessible to environmental scientists through open-source software. We chose to experiment on datasets containing 2000 labelled points per class, but when randomly subsampling the labelled data, we observed that high accuracies are already possible with a few hundred ground truth points per class, as featured in Table 6. Neural networks are also more abstract and thus harder to decipher, contrary to 3DMASC, thanks to feature importance and Shapley values that contribute to an explainable machine learning approach. Indeed, we refer to the definition of Roscher et al., 2020, identifying transparency, interpretability, and explainability as major traits of explainable machine learning. In this work, transparency is addressed through the detailed description of the method construction choices and the exploration of some key parameters such as feature selection choices, scale choices, etc. Interpretability is addressed through extensive analysis of the features' importance, the iterative pruning of the descriptor vector, the exploration of meaningful scales and the consideration of two different environments and datasets. Finally, explainability is considered through the analysis of the decision process in the two environments depending on the class – typical features and scales for different classes – but also through the experiments on two datasets and the application to much larger sets of points, giving insights on the robustness and reproducibility of the method.

| Samples per class | Overall accuracy | | | |
|---|---|---|---|---|
| | AIN | | FREHEL | |
| | 5 cl. | 11 cl. | 5 cl. | 13 cl. |
| **1600** | 98% | 95% | 91% | 91% |
| **1200** | 98% | 95% | 91% | 91% |
| **800** | 97% | 95% | 91% | 90% |
| **400** | 96% | 95% | 90% | 90% |
| **100** | 94% | 93% | 89% | 90% |

Table 6: Classification accuracy depending on the number of training samples used. Tests are performed using the complete set of 3DMASC features.

**6.2 Dominant scales**

Taking advantage of the explainability of the method, we identify typical characteristics of OC classifications. First, **a typical set of scales emerges from the experiments**, including small and medium sphere diameters ranging between 1.5 m and 7.5 m and one larger scale around 14 m (Table 3). The global range of scales selected does not vary between primary and advanced classifiers, except for the Ain, where we can expect that the introduction of smaller-scale objects - *vehicles, swimming pools, intermediate vegetation* – penalizes very large scales. Advanced classifiers rather add scales within the core range, reducing the step between two options. Second, **the exact optimized sets of scales that arise are specific to each environment**, which questions the possibility of identifying optimal neighbourhoods without analyzing their application context. For example, out of four experiments, three different optimized NN neighbourhoods stand out: 10, five, and one (see Table 3), contrasting with the conclusions of Niemeyer et al. (2011) that select one single optimal scale of seven NN for their different experiments, and with the results of Dong et al. (2017) who also find a single scale of five NN as the most often selected neighbourhood.

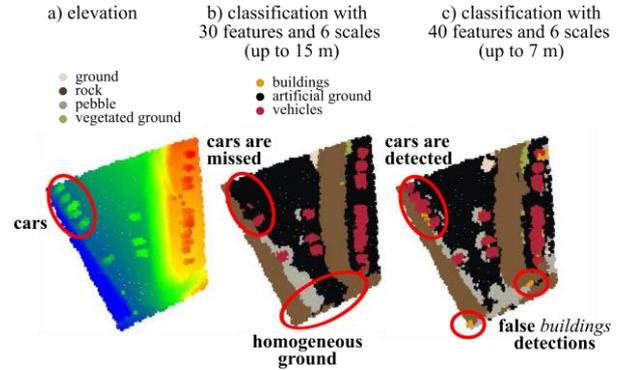

Figure 13: Extracts of classification results obtained depending on the maximal scale included.

Furthermore, the fact that each selected scale is not used for each feature tends to be consistent with the work of Dong et al. (2017), choosing to optimize each feature's neighbourhood rather than identifying a global optimal scale. Third, **scale selection results are consistent with the intra-feature correlations** we observed in Figure 5. Although these estimates could be complemented with other measurements of non-linear correlations, as is made in (Weinmann et al., 2015), this first consideration for correlation already provides insightful information on the relevance of the different scales for our study areas. These suggested that fewer scales were needed above 6 m than below, which is in line with the fact that we only obtained one large scale. This large scale also outlines the necessary **trade-off between classification accuracy and classification resolution**. If we investigate the role of this much larger scale, we find that, though it helps to mitigate some errors linked to larger scale roughness in the PCs – for example, confusion of rocks with buildings – it also smoothes out the results, blurring classes borders and even missing smaller objects like cars. In Figure 13, cars can be identified in the PC, but many of them are missed and labelled as ground when large scales are used. Limiting the range of scales to 7 m produces a result in which these cars are correctly detected, but the ground incorporates false *building* labels.

Our observations thus **question the relevance of large scales**, which appear to be selected for certain point types as they pass the score filtering selection but end up penalizing the global classifier application through several aspects. Table 7 illustrates the confidence filtering analysis obtained on the Fréhel advanced classifier optimized on scales within 1 to 7 m only. It shows that, without the possibility to select larger scales, the classification reaches similar accuracies and confidences. However, they clearly affect the computation efficiency. The advanced Fréhel OC classifier obtained on scales up to 7 m incorporates ten more features, but the computation time is divided by three (3450 points per second versus 1102 points per second). Suppressing large scales may thus improve classification speed while maintaining high OAs.

| Classifier | Confidence threshold | OA | Remaining points (%) |
|---|---|---|---|
| **FREHEL advanced (Max scale = 7 m)** | 0.5 | 94% | 91% |
| | 0.6 | 96% | 84% |
| | 0.7 | 97% | 76% |
| | 0.8 | 98% | 66% |

Table 7: Overall accuracy depending on the confidence threshold for a reduced set of possible scales.

**6.3 Computation time**

Computational efficiency is an important aspect of 3DMASC. The computation of the spherical neighbourhoods is the main bottleneck of the workflow, similar to what is observed in other studies (Hackel et al., 2016; M. Weinmann et al., 2015), and sometimes even drives the choice of the neighbourhood type. Table 8 illustrates the time necessary to compute all implemented features at different single or combined scales. These results were obtained using a computer equipped with 128 Go of memory and a 12-core AMD Ryzen ™ 9 5900X CPU. The test file was the data of the Ain area; it contained 106 410 018 green points, 61 043 388 NIR points, and 5 700 844 core points having a 1 m spacing. Table 8 shows how crucial scale selection and optimization are: without optimization, computing scales from 1 m to 10 m lasts 5 hours and 45 minutes (275 pts/s). After optimisation using scales up to 7 m, the computing time drops to ~28 minutes (3450 pts/s). This computation speed could be increased by implementing pyramidal computation into the 3DMASC plugin, which consists of subsampling the data when increasing the neighbourhood size, as made in Thomas et al., (2018). Additionally to the selection of a scale range, the number of different diameters within the interval and the number of features to compute for each neighbourhood also has an impact – though less significant – on the processing time. Table 8 shows that the optimised descriptor set relying on scales up to 7 m is three times faster to compute than the complete set of features on scales between 1 and 7 m. Consequently, although predictor selection is not crucial for classification performance (see Table 3), it is essential to the practical applicability of the method.

**6.3 Class-wise results: dominant features**

**Five of the maximal 12 features needed to perform basic classification are common to both experiments**. These are *multi-echo features computed on both PCs, vertical distance of green points to their NIR neighbours (kNN), and mean green intensity.* They are then combined with site-specific attributes. In Fréhel, the optimized predictor set retains mainly multi-echo attributes and intensity-derived information. In the Ain, multi-echo features and height-derived parameters dominate. However, **classical features of 3D data interpretation,** such as dimensionality-based features delineating the shape of local PCs from the combination of eigenvalues (Brodu and Lague, 2012; Gross and Thoennessen, 2006; Vandapel et al., 2004; Weinmann et al., 2013) **are almost unused**. They only become more prominent when complexifying the number and types of classes to detect. This is also certainly linked to the fact that we analyze airborne lidar data, while these features were designed in priority to describe terrestrial and mobile laser scannings that include a greater diversity of surface orientations. **Point-based attributes are also absent** from the optimized classifications. Newly introduced features based on statistical operators applied to multi-echo features or intensity values systematically outperform them in terms of contribution. Such operators had been tested on height-derived values (Antonarakis et al., 2008; Dong et al., 2017) but never applied to other types of features. **The use of statistical operators is particularly informative and able to drastically improve the informative power of point-based characteristics**, namely multi-echo attributes, that never particularly stood out in existing PC classification literature but appear essential to the success of our experiments. We interpret this result as a way for decision trees to compensate for their inability to consider spatial relationships between points and include a level of spatial consistency of the considered attributes. These **operators also limit bias linked to intensity values, which are unavoidable** in classifying diverse environments (Song et al., 2002; Yan et al., 2015). Intensity median, mode, skewness, or ratio values constitute half of the primary predictors in Fréhel and are prominent in both advanced models. By being less sensitive to outliers, standard deviation and skewness mitigate the limitations of this measure, which varies with the acquisition conditions and does not constitute an absolute estimation of surface reflectance (Kashani et al., 2015). Overall, **the features we present seem to describe natural environments better**, as the results of importance-based feature selection illustrate. To further support this observation, an ablation study made on the Ain River is presented in Supplementary Materials, and shows that weaker results are obtained without the 3DMASC features in terms of accuracy, generalisability, and classifier complexity.

|  | 1 m | 4 m | 7 m | 10 m |
|---|---|---|---|---|
| **Single scale (pts/s)** | 28082 | 7337 | 2109 | 928 |
|  |  | 1 – 4 m | 1 – 7 m | 1 – 10 m |
| **Multi-scale (pts/s)** |  | 4069 | 854 | 275 |
|  |  |  | Up to 7 m | Up to 15 m |
| **Optimised multi-scale (pts/s)** |  |  | 3450 | 1102 |

Table 8: Feature computation time depending on the scale set.

We compared classifications of the Ain obtained with 3DMASC features and with features used in Thomas et al. (2018), Hackel et al. (2016), Chehata et al. (2009), and Rusu et al. (2009). In order to ensure a fair comparison, all features were computed at the same scales and on the green PC only and then classified with a RF model. Consequently, the 3DMASC version that is compared to these existing methods only includes single-cloud features. In practice, only the features differ, with 3DMASC incorporating spatial statistics of multi-echo attributes and all other features but no information about the NIR cloud – and thus no dual cloud feature or contextual feature. Additionally to the point feature histograms developed by Rusu et al. (2009), the compared approaches rely mainly on features derived from the covariance matrix of the core point's neighbourhoods, height-based parameters, and, less frequently, echo-based parameters. Due to the unavailability of waveform data in our test areas, we omitted waveform-derived attributes initially exploited in Chehata et al. (2009). We only computed each feature set on the green PC and at multi-scale spherical neighbourhoods with diameters of 2, 3, 4, 5, 6, and 7 m. The feature point histograms were computed only on the green PC, with a normal computation scale of 0.5 m and a feature computation scale of 2 m. Histogram-based features also involve the spatial repartition of features (Osada et al., 2002; Rusu et al., 2009; Tombari et al., 2010). Although they have been used for classification before with satisfactory results (Arbeiter et al., 2012; Blomley et al., 2016; Blomley and Weinmann, 2017; Garstka and Peters, 2016; Himmelsbach et al., 2009; Wohlkinger and Vincze, 2011), we made a choice not to include them in 3DMASC to avoid increasing the predictor selection difficulty, as they require the choice of two scales – one to compute the normal, the other to compute the features

| Dual cloud | Single cloud | | | | |
|---|---|---|---|---|---|
| Dual cloud 3DMASC | Single cloud 3DMASC | **Thomas et al. (2018)** (covariance- based) | **Hackel et al. (2016)** (covariance- and height-based) | **Chehata et al. (2009)** (covariance-, height-, echo-, plane-based) | **Rusu et al. (2009)** (fast point feature histogram) |
| **97.6%** | 93.7% | 74.3% | 82.6% | 84.9% | 71.3% |
| **Combined with single cloud 3DMASC** | | 93.7% | 93.7% | 93.7% | 93.7% |

Table 9: Classification overall accuracies obtained with different types of single-cloud features on the 5 classes of the Ain dataset

– and selecting the number of bins to use. Their higher computation time also affected this decision (Garstka and Peters, 2016). Details about the features used in each experiment are provided in Supplementary Materials. Overall Accuracies obtained on the test set for five classes in the riverine area by each approach are summed up in Table 9. They show that in natural environments, using our features produces systematically higher results than other existing features and that none of these features, particularly the fast point feature histogram, improved the single cloud 3DMASC classification results. They also show that using solely covariance-based features produces OAs among the lowest in our riverine environment, while it generates more precise classifications of urban environments (Thomas et al., 2018), highlighting the need for methods adapted to the different types of 3D data currently in use.

We also **introduce new measures of the optical behaviour of the surfaces present** in the PCs, which were mostly estimated through mean intensity, and propose new inter-channel ratios to complement existing multispectral attributes (Morsy et al., 2017b; Wichmann et al., 2015). Previous studies analyzing multispectral lidar faced the difficulty of linking points to their equivalents in PCs of other wavelengths since they are never in strictly identic positions due to the sensor configuration (Lague and Feldmann, 2020). These new ratios, along with our dual-cloud features, compensate for the limits of point matching used in existing multispectral lidar analysis work (Morsy et al., 2017b) when they are used on datasets with correct geometrical and radiometric calibration (Kashani et al., 2015; Yan et al., 2012). **Dual-cloud features systematically stand out among highly contributive features**. Their lower inter-scale correlation likely contributes to their more informative character, along with their ability to compensate for the limits of shallow learning classifiers that cannot learn features and thus to bring out and use connections between features. For example, the difference of roughness between the NIR and the green PC is particularly high for points belonging to the water column and much lower for the riverbed or the bottom of swimming pools due to the full reflection of the NIR laser on the water surface and the scattering of the green light in the water column. The same optical phenomenon explains the higher difference in elevation modes between PCs in swimming pools and rivers. The inherent point position differences of TB sensors, illustrated in Figure 1, explain the varying vertical distances between green and NIR PCs in vegetated areas and their systematically negative value in bathymetric zones. Similarly**, the use of a previously classified ground PC as a contextual feature** allows for improvement in the labelling of points at the limit between ground and above-ground features, namely building walls and lower tree branches, explaining the improvement observed when they are included and the smaller scales needed to capture the signature of such variations.

Using these observations, we **recommend the following set of features to use on topo-bathymetric environments**: the NIR and green *number of returns* and *echo ratios*, the green *return number*, the *vertical distance* to the 1 and 10 *nearest neighbours of the core points in the NIR PC*, the *mode* of the green and NIR intensities, the *skewness* of the NIR *intensity*, the *ratio of median intensities*, the NIR and green *elevations' standard deviation*, the *difference of elevation modes*, the NIR *roughness* and the *difference of roughness*, the NIR *dip*, and the green PC *sphericity*. With these 19 features computed at scales between 1.5 and 14 m, we observed OAs of 98% and 91% for five classes on the riverine and coastal datasets, respectively, and 94% and 90% on their 11 and 13-class versions.

# 7. CONCLUSION

In this paper, we have introduced 3DMASC, a method for explainable machine learning multispectral point cloud classification. 3DMASC operates directly on sets of unordered, unstructured points and predicts a label for each, with a confidence index and information on the origin of the decision through feature importance. It differs from previous point cloud classification methods in its capacity to handle multiple clouds simultaneously and describe the spatial and statistical repartition of point cloud attributes, introducing indirect context consideration in the model and new multispectral feature ratios. 3DMASC also stands out from state-of-the-art 3D classification methods with its accessibility: it is explainable using Shapley values, usable without dedicated GPUs, and easy to handle for thematic specialists such as geomorphologists, ecologists, or cartographers. We focus on providing compromise in terms of computation cost, processing time, complexity, and resulting metrics with respect to the current state-of-the-art methods. We demonstrate the performance of the approach on two different airborne lidar use cases: the detection of land and sea covers in (1) a fluvial environment and (2) a coastal area. Results show that the method produces highly accurate classifications of basic or detailed categories of points. Furthermore, models excel in TB environments thanks to the newly introduced features and require limited training points ($\leq 2000$ per class), scales, and attributes. We also implemented a feature selection framework that allows us to draw three main conclusions about the definition of the predictor's vector: (1) statistics of point-based attributes are more informative than classical dimensionality or geometrical features on this type of data, (2) multi-echo features, vertical distance between the two PCs and mean intensity appear to constitute an essential base of features to use and (3) dual cloud features are highly contributive to separate ground, artificial elements and vegetation. Our results also stress multi-cloud classification's superiority over single-cloud, especially for bi-spectral lidar. We release our source code through an open-source plugin in

CloudCompare (Girardeau-Montaut, 2022), hoping it will help applications of 3D remote sensing for earth observation and conservation. Although our paper illustrates specific use cases of the workflow on topo-bathymetric lidar datasets, 3DMASC can be extended to PC time series analysis and 3D data interpretation in general. It may be applied to terrestrial laser scanning data, to structure from motion PCs, or even to data acquired with drone lidar sensors, which are still too compact to incorporate dual-wavelength lidar sensors but already enlarge the access to lidar surveys.

**Acknowledgements:** This research was partially funded by the Saur Group and the Region Bretagne, whom the authors thank for their support. The Titan DW sensor, operated by the Nantes-Rennes Lidar Platform has been funded by the Region Pays de la Loire with funding of the RS2E-OSUNA programs and the Region Bretagne with support from the European Regional Development Fund. Patrick Launeau is greatly acknowledged for his contribution in the acquisition of the Titan DW sensor. We thank Cyril Michon, Emmanuel Gouraud, William Gentile from Geofit-Expert company, and Laurence Hubert-Moy for their contribution to the overall operation of the Titan DW sensor. We thank Electricité De France (A. Barillier, A. Clutier) for commissioning the acquisition of the Ain River survey and providing access to the data.

**Appendix A: Cloudcompare (CC) q3DMASC plugin implementation and operation**

Using the q3DMASC plugin for classifier training or inference requires a labelled core point file and up to 3 accessory point clouds used to compute the features around each core point: PC1 (e.g., green channel), PC2 (e.g., NIR channel) and CTX (e.g., a point cloud with a populated classification field). For single-point cloud classification, only one accessory point cloud is needed. A text file contains the description of point clouds, scales, and features to be used for training. Upon training completion, a classifier file is saved and can be subsequently used with q3DMASC to apply the classifier to other point clouds.

Here are the main characteristics of the q3DMASC plugin implemented in the open-source software CloudCompare (CC):

**Accessibility**: the q3DMASC plugin has been designed to be usable without programming language knowledge (e.g., Python) directly in the CC GUI. As such it makes a great introductory tool for non-specialists, for teaching and for quick tests without having to setup a complete programming environment. We have also modified the CC scissor tool to allow direct interactive labelling of 3D data and introduced a tool to split point clouds according to classes automatically, and a new plugin for labelling data in 3D has just been released (QCloudLayers by Wiggins Tech). These simple tools associated with the neat 3D visualization of CC greatly facilitate the creation of labelled 3D data for training.

**Speed:** (CC) written in C++ has a well-proven, fast and fully parallelized 3D neighbourhood search essential for fast computation of spherical neighbourhood or kNN search. While not critical during the training phase as a limited number of samples is necessary, this is essential during application and production phases to compute features on several millions of points without requiring a specific configuration (e.g. GPU).

**Scalability:** the q3DMASC plugin can be used in command line mode without GUI in order to apply the classifier in batch mode for large point cloud projects that would not fit in the computer memory. For instance, we have been able to use it routinely to process projects with more than 10 billion points using tiling strategies.

**Non data source specific:** while some features of 3DMASC are specific to Airborne lidar (e.g., multi-echo features), many geometric features can be used for any high-resolution 3D point cloud created, for instance, from terrestrial lidar, Structure From Motion (SFM), Satellite Stereo Photogrammetry and multibeam sonar. There are, in particular, provisions to use RGBNIR information that can be essential for SFM.

**Flexibility in feature creation**: to generate complex single or dual cloud features over several scales, the user has to create a text file containing the description of the various point clouds, the scales to be used, and the features to be computed. Complex single cloud features can be generated using the following formalism:

$$FEAT\_SC\#\_STAT\_PC\#,$$

in which FEAT corresponds to a predefined list of features (e.g., *intensity, z, number of returns, sphericity, …*), SC# indicates the scale at which they will be calculated, *STAT* is a statistical descriptor for point-based features sampled within the spherical neighbourhood (*mean, mode, median, std, range, skew*), PC# indicates the point cloud to be used for calculation around the core point. Dual cloud features are generated with this formalism:

$$FEAT\_SC\#\_STAT\_PC\#\_PC\$\_MATH$$

In which PC$ indicates the second cloud to be used, and *MATH* is an operator (*minus, plus, divide, multiply*). For instance, the Z mode difference (Figure 2) between the green channel (PC1) and the NIR channel (PC2) calculated at all possible scales is written Z_SCx_MODE_PC1_PC2_MINUS. Contextual features are constructed using the following formalism:

$$DZk\_SC0\_PC\#\_CTX\#,$$

In which DZ$k$ (resp. DH$k$) indicates the vertical (resp. horizontal) distance to the $k$ nearest neighbours, PC# indicates the PC considered, and CTX# is the number in the classification field to consider (e.g., 2 for ground, 5 for vegetation…). For instance, the average vertical distance to the three nearest ground points of the NIR channel (PC2) that holds a valid classification field is DZ3_SC0_PC2_CTX2.

**Explainability**: we use a random forest algorithm that combines a good performance on many attributes, simplified feature selection, and robustness to overfitting. After training, the GUI version of 3DMASC outputs the overall accuracy and RF feature ranking and allows the manual removal of less contributing features. After training completion, users can directly visualize feature values in 3D to understand why they contribute directly or not to classification success.

For training purposes, we chose the cross-platform OpenCV library (Bradski, 2000) implementation of Random Forests as it allows classifiers created in Cloudcompare to be used in Python and vice-versa. The downside of the C++ implementation of OpenCV is that the training is not parallelized and is consequently much slower than the RF implementation, e.g., of scikit-learn (Pedregosa et al., 2011). RF training is thus the main bottleneck during classifier creation in the CC version. Classifier application is extremely fast, and feature calculation becomes the main bottleneck. Expert users can directly train their classifier in Python with their favourite algorithm.

**Appendix B: Complete list of features used in this study**

Point-based features and single/dual cloud features constructed from them in spherical neighbourhoods:

| Name | Single cloud features stat descriptors (532 nm or 1064 nm) | Dual cloud features (532 and 1064 nm) | |
|---|---|---|---|
| | | Subtraction | Division |
| Elevation* | Std, Skew | Mean, Median, Mode, Std, Skew | - |
| Intensity | X | Std, Skew | Mean, Median, Mode |
| Return number | Mean | - | - |
| Numb. of returns | Mean | - | - |
| Echo Ratio | Mean | - | - |
| R, G, B | Mean, Mode, Median | - | - |

*: not used as a point-based feature

Dimensionality-based features computed in spherical neighbourhood:

| Name | Formulation from eigenvalues | Dual cloud features (532 and 1064 nm) |
|---|---|---|
| PCA1* | $\lambda_1/(\lambda_1+\lambda_2+\lambda_3)$ | subtraction |
| PCA2* | $\lambda_2/(\lambda_1+\lambda_2+\lambda_3)$ | subtraction |
| PCA3/Surf variation+ | $\lambda_3/(\lambda_1+\lambda_2+\lambda_3)$ | subtraction |
| Sphericity+ | $\lambda_3/\lambda_1$ | subtraction |
| Linearity+ | $(\lambda_1-\lambda_2)/\lambda_1$ | subtraction |
| Planarity+ | $(\lambda_2-\lambda_3)/\lambda_1$ | subtraction |

*: Brodu and Lague (2012); + : Weinmann et al., (2013)

Geometry-based features computed in spherical neighbourhood:

| Name | Information | Dual cloud features (532 and 1064 nm) |
|---|---|---|
| Verticality* | Varies between 0 (horizontal) and 1 (vertical) | subtraction |
| Detrended Roughness | Std of distance between points and best fitting plane | subtraction |
| Curvature | Mean curvature in CC= average of principal curvatures | subtraction |
| Nb of points | - | subtraction |
| Anisotropy | Ratio of distance to center of mass and radius of sphere | subtraction |
| First Order Moment* | Hackel et al. (2016) | subtraction |

*: Demantké et al., 2012;

Height-based metrics computed in spherical neighbourhood:

| Name | Formulation | Dual cloud features (532 and 1064 nm) |
|---|---|---|
| Zrange | $z_{max}-z_{min}$ | subtraction |
| Zmin | $z-z_{min}$ | subtraction |
| Zmax | $z_{max}-z$ | subtraction |

$z$ is the core point elevation, $z_{max}$ and $z_{min}$ are the maximum and minimum elevation in the spherical neighbourhood, respectively.

Contextual features in the NIR channel:

| Name | Formulation | Target class |
|---|---|---|
| DZ to kNN | Mean vertical distance to k nearest neighbour | 1064 nm ground |
| DH to kNN | Mean horizontal distance to k nearest neighbour | 1064 nm ground |

**Supplementary materials :**

Figure 1 presents the results of the experiments made to determine the scale to use for evaluation and the correlation threshold to apply during feature selection (OA=Overall Accuracy).

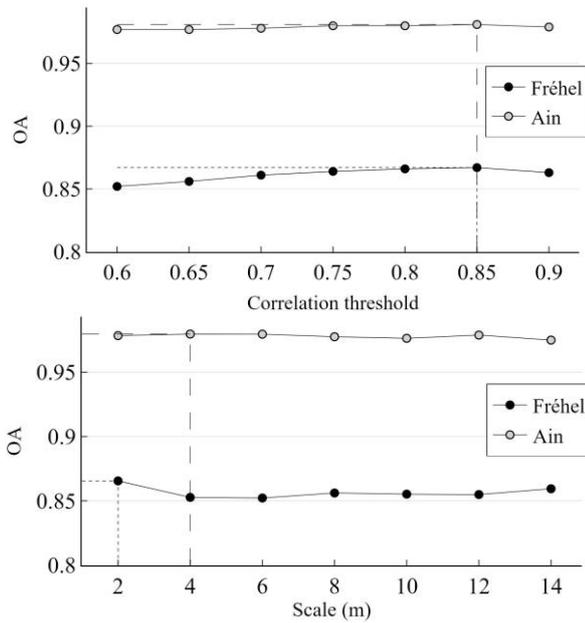

Figure 1: Results of the experiments performed to determine the correlation threshold to apply for feature selection and the scale at which to evaluate each feature.

Figure 2 illustrates the variation of the OA depending on the number of predictors used. The predictors used at each iteration are identic to those involved in the realization of Figure 7 in the paper. This figure shows that OA and OOB display similar dynamics when pruning the predictors set.

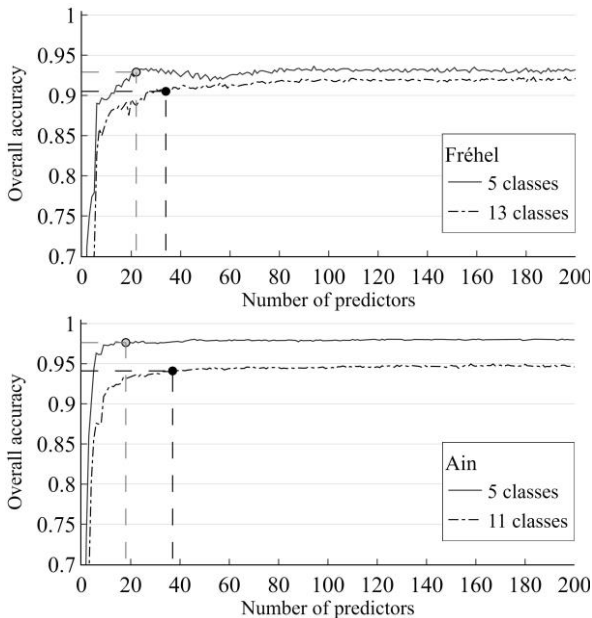

Figure 2: Overall accuracy depending on the number of predictors used for each experiment.

Figure 3 gives more detailed information on the correlation between features computed at different scales.

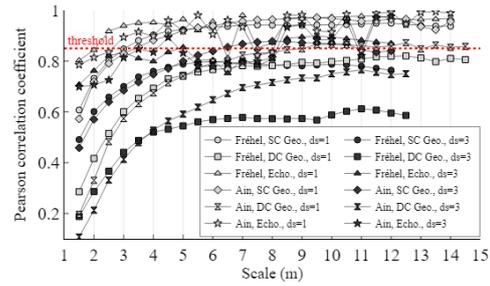

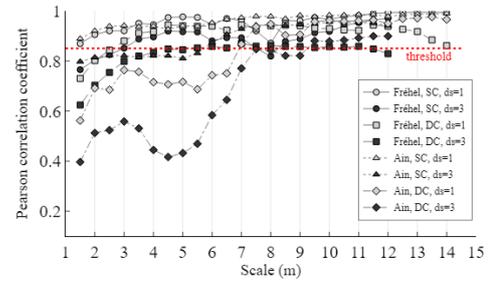

Figure 3: Linear correlation between features computed at scales separated by ds=1 m or ds=3 m for different families of features. SC = Single Cloud ; DC = Dual Cloud

Figure 4 illustrates the impact of large scales on classification accuracy depending on the presence or absence of contextual features (vertical distances to a previously classified ground point cloud). Scales were removed iteratively per decreasing order.

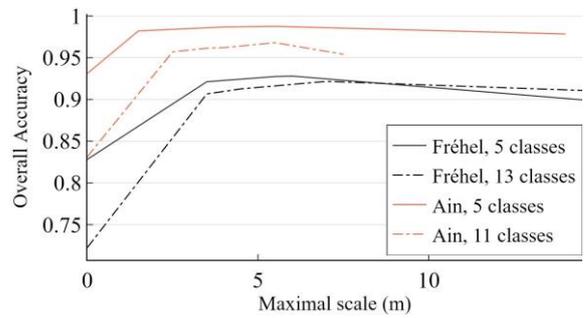

Figure 4: Classification performances depending on the maximal scale kept in the optimized predictor set and in the predictor set augmented with contextual attributes.

Table 1 details the features used to compare 3DMASC to other approaches. The eigenvalues referred to are those obtained on the covariance matrix of spherical neighbourhoods. For detailed mathematical expressions of the different attributes, please consult the original papers (Chehata et al., 2009; Hackel et al., 2016; Thomas et al., 2018). The comparison to the fast point feature histogram introduced in Rusu et al. (2009) was made using the dedicated computation framework implemented in the Python Open3D library (Zhou et al., 2018).

| Name | Thomas et al. (2018) | Hackel et al. (2016) | Chehata et al. (2009) |
|---|---|---|---|
| Sum of eigenvalues | X | X | |
| Omnivariance | X | X | |
| Eigenentropy | X | X | |
| Anisotropy | | X | X |

| Feature | | | |
|---|---|---|---|
| Linearity | X | X | X |
| Planarity | X | X | X |
| Sphericity | X | X | X |
| Curvature | | | X |
| Surface variation | X | X | |
| Verticality | | X | |
| Verticality based on 1st eigenvector | X | | |
| Verticality based on 3rd eigenvector | X | | |
| Vertical moment (1st order) | X | | |
| Vertical moment (2nd order) | X | | |
| Number of points | X | | |
| Statistical moments of eigenvectors (1st and 2nd order) | X | X | |
| Z range in neighbourhood | | X | |
| Difference with minimal Z in neighbourhood | | X | X |
| Difference with maximal Z in neighbourhood | | X | |
| Standard deviation of Z in neighbourhood | | | X |
| Residuals of the fitting of a plane to the neighbourhood | | | X |
| Deviation angle of a fitted plan normal to the vertical | | | X |
| Variance of the deviation angles of the three dimensions of the neighbourhood | | | X |
| Distance to the fitted plan | | | X |
| Number of returns | | | X |
| Normalised return number | | | X |

Table 1: description of the features used in each approach compared to 3DMASC on the Ain dataset.

Figure 5 shows the out-of-bag scores obtained for the five-class classification of the Ain River depending on the number of predictors during classifier optimization. Two cases are presented: one in which the initial set of features includes the newly introduced 3DMASC features, and another in which they are not included. The resulting curves illustrate the lower generalisability obtained without the 3DMASC features – which produce a globally lower OOB score – and the higher number of predictors required for the OOB score to converge – which denotes the necessity to use more predictors to get a stable classifier.

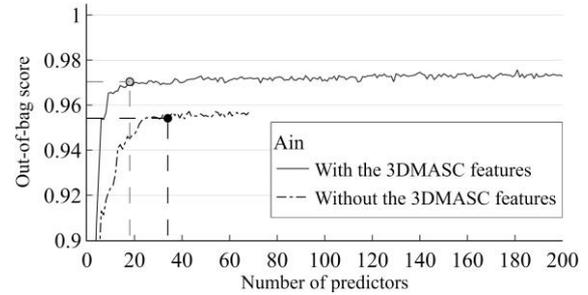

Figure 5: Out-of-bag score for the five-class classification of the Ain River depending on the inclusion of 3DMASC features in the set of predictors and on the number of predictors used for each experiment.

In Figure 6, the evolution of the overall accuracy during the optimization process depending on the inclusion of the 3DMASC features in the initial feature set is presented. Globally, both Figure 4 and 5 suggest that 3DMASC features outperform classical features in terms of informative power.

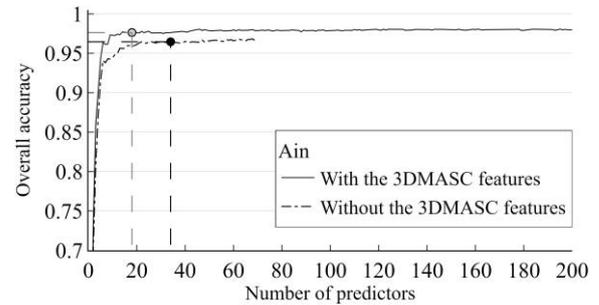

Figure 5: Overall accuracy for the five-class classification of the Ain River depending on the inclusion of 3DMASC features in the set of predictors and on the number of predictors used for each experiment.